# Micromechanical characterisation of osteoarthritic subchondral bone by micropillar compression

Samuel McPhee[1], Marta Peña Fernández[1], Lekha Koria[2], Marlène Mengoni[2], Rainer J Beck[3], Jonathan D Shephard[3], Claire Brockett[4], Uwe Wolfram[1,5*]

[1] Institute of Mechanical, Process and Energy Engineering, Heriot-Watt University, UK.
[2] Institute of Medical and Biological Engineering, University of Leeds, UK.
[3] Institute of Photonics and Quantum Sciences, Heriot-Watt University, UK.
[4] INSIGNEO Institute for in silico Medicine, University of Sheffield, UK.
[5] Institute for Material Science and Engineering, TU Clausthal, Germany.
* Corresponding author: uwe.wolfram@tu-clausthal.de (Uwe Wolfram)

## Abstract

Osteoarthritis (OA) is a multifaceted joint disease which poses significant socioeconomic burdens and remains a significant clinical challenge. Evidence suggests that structural and mechanical changes in subchondral bone influence the pathogenesis and development of OA, leading to diminished bone quality and cartilage degeneration. While changes in microstructure and tissue scale elastic properties are well reported, the tissue yield response of subchondral bone in OA and their correlation with compositional changes have not been investigated. Here, we performed quasistatic micropillar compression and nanoindentation within the subchondral bone plate and trabeculae of hydrated non-diseased (ND) and OA affected specimens retrieved from the distal tibia *in vivo*. The micropillars, extracted by laser ablation, exhibited a taper angle which mandated the use of an *in silico* micropillar compression routine to back-calculate elastic modulus and strength of the bone tissue that comprised each micropillar. Elastic modulus remained unchanged between ND and OA subchondral bone, whereas strength increased from 46.0 MPa to 57.3 MPa in OA subchondral trabecular bone but not in the bone plate. Micropillar matched Raman spectroscopy and quantitative backscattered electron imaging revealed mineralisation is the underlying determinant of elastic modulus and strength at the microscale. By combining micromechanical and tissue compositional analyses, we investigated how the mechanical properties are related and how these properties are affected in subchondral bone by OA. Our results may be of value in the development and optimisation of interventions used to alleviate the socioeconomic burdens associated with this debilitating joint disease.

**Keywords:** Osteoarthritis, Subchondral bone, Micropillar compression, Strength, Raman spectroscopy, Quantitative backscattered electron imaging

## Statement of significance

Osteoarthritis imposes significant socioeconomic burden, necessitating ever improving treatment strategies. Abnormal mechanical forces at the joint, such as injury, overloading, or aging are associated with osteoarthritis development. While healthy tissue is well understood, the impact of diseases on tissue mechanical properties needs to be brought on a similar level of understanding. We present in this study the first use of micropillar compression for assessing the tissue scale compressive strength of subchondral bone in osteoarthritis. Strength was increased in osteoarthritis compared to non-diseased tissue only in trabeculae subjacent to the bone plate. Compositional analysis confirmed mineralisation as a key determinant of tissue mechanical properties. This integration of micromechanical and compositional analyses sheds light on tissue properties and how these are affected by osteoarthritis in subchondral bone.

# 1. Introduction

Osteoarthritis (OA) is a multifaceted joint disease that significantly impacts the quality of life for those affected. It causes chronic pain, diminished mobility, and a decline in mental well-being [1–3]. Globally, over 500 million people had OA in 2020, and its incidence is predicted to continue to increase as populations age [4], leading to significant economic, social, and health burdens worldwide. The heterogeneous pathophysiology of OA limits the efficacy of pharmacological interventions and there are currently no approved disease modifying OA drugs which can prevent or slow the progression of the disease [5]. Early-to-mid stage symptomatic OA can be managed with a mixture of patient education, structured exercise and anti-inflammatory pharmaceuticals [6]. However, the efficacy of such methods decline as the disease progresses and the focus turns to surgical interventions to deliver satisfactory clinical outcomes for late and end-stage OA [7].

Taking ankle OA as an example, surgical interventions include osteotomy, arthrodesis, and arthroplasty [8]. The former usually involves surgical realignment of the joint through a wedge opening, most commonly at the distal tibia which is stabilised by a plate [9]. The resultant realignment redistributes load across the joint to alleviate symptoms and slow down OA progression [8,9]. At later stages, fusion of the joint (i.e., arthrodesis) through fixation of the distal tibia and talus into an aligned neutral position has shown great success in relieving pain, albeit inhibiting ankle motion [10]. An alternative to improve ankle mobility is total ankle arthroplasty, where the ankle joint is replaced with a prosthetic device, however, the survival rates of these prostheses are lower than those used for other joints [11,12].

Regardless of the stage, mechanical loading plays an essential role on the pathogenesis, progression, and treatment choice for OA [13,14]. On the one side, OA manifests in highly stressed regions in the joint as a result of overloading, injury, or aging [14]. This pathological mechanical environment is associated with alterations to subchondral bone remodelling which may alter subchondral and trabecular bone morphology [15], and stiffness [16] leading to increased shear stresses in the cartilage and consequent degeneration and OA development [14]. On the other side, surgical treatments aim to restore the natural biomechanics of the joint, through the insertion of implant devices [13]. However, the loss of bone quality due to OA may impair the stabilisation and osseointegration of such implants, leading to premature failure and early need for revision. A comprehensive material characterisation of the osteochondral unit, and in particular, the underlying subchondral bone tissue is crucial for improving current treatment of OA [14].

Some micro- and nanoindentation studies have investigated the impact of OA on subchondral bone, reaching a consensus that tissue scale elasticity increases with disease progression [16–18]. Beyond evaluation of the tissue scale elastic response of OA subchondral bone, to the best of our knowledge, there are no studies which explicitly investigate tissue scale strength. Consequently, our primary objective was to investigate both the stiffness and yield-strength of subchondral bone at the tissue level in OA and their correlation with compositional changes.

To facilitate this investigation, we employ micropillar compression, which like nanoindentation is principally an indentation technique. Unlike nanoindentation, micropillar compression allows us to facilitate a compression test which, in turn, allows us to determine strength. Within the remit of sample preparation, micropillar compression requires the additional step of fabricating micrometre sized cylinders, or pillars. These micropillars are then compressed by a flat probe, inducing a homogenous uniaxial stress state within the micropillar. The recorded force-displacement data can then be transformed to a uniaxial stress-strain relationship provided that the micropillar geometry is known. Micropillar compression has been used to investigate the influence of tissue hydration [19,20], osteogenesis imperfecta [21], aging [22], strain-rate dependency [23], and lamella orientation [19,24] on the post-yield behaviour of cortical bone at the microscale. The challenge posed by fabricating micropillars potentially limits the accessibility of the technique. The extraction process requires the removal of an annulus shaped channel to isolate a pillar that protrudes from the bulk tissue. Early adopters of the technique for testing bone used focussed ion beam milling as a means to extract micropillars [19,25,26] However, this technique is expensive and time-intensive. Several authors sought to overcome these constraints and advocated the use of ultrashort pulsed laser ablation for extracting micropillars which enabled higher-throughput testing [22,26–28].

We adopt their methodological concept and aim to investigate micromechanical and compositional changes in OA subchondral bone. Specifically, we: (i) Fabricate micropillars directly by ultrashort pulsed laser ablation, located in the subchondral bone plate and subjacent trabeculae of non-diseased and late-stage ankle OA bone; (ii) Conduct nanoindentation and quasi-static micropillar compression to evaluate microscale elastic and yield strength of rehydrated subchondral bone; (iii) Use Raman spectroscopy and quantitative backscattered electron microscopy to evaluate relative changes to tissue composition within each micropillar.

# 2. Materials and Methods

## 2.1. Sample preparation

Osteoarthritic (OA) tibial specimens were sourced *in vivo* from two patients (males, ages 65 and 68 years) undergoing total ankle arthroplasty. Both patients exhibited radiographic features of grade 4 OA (Kellgren-Lawrence scale) [29]. Two sections were cut from the provided bone samples from each patient. Ethics approval was granted by the University of Leeds MEEC research ethics committee (MEEC 18-027) and NHS Yorkshire and Humberside National Research Ethics Committee (REC 07/Q1205/27) for the ND and OA specimens respectively. Non-diseased (ND) tibial specimens were collected from three cadavers (males, ages 43, 50 and 57 years). A distal tibia was extracted from each cadaver (n=3), cleaned of soft tissue, and further dissected by diamond-blade bandsaw (Exakt, Germany) under constant water irrigation. Cuts were made along the central sagittal plane to extract two sections from the tibial plafond.

We followed previous embedding and polishing protocols for performing micro- and nanoindentation of bone specimens [30,31]. In short, the sections were first cleaned by water irrigation and ultrasound bath exposure before air drying for 24 hours. We then embedded the sections in PMMA (Technovit 4006SE, Heraeus, Germany), with an initial exposure to an elevated pressure of 2 bar for 30 minutes to promote PMMA infiltration. After at least 24 hours of curing, the samples were parallelised and polished using progressive grades of silicon carbide paper (P800, P1200, P2000, P4000) and finished with an 0.25 µm diamond suspension (Hermes Abrasives, Germany).

## 2.2. Micropillar extraction

We extracted micropillars using ultrashort pulsed laser ablation. To do so, we modified our previous micromachining protocols used to extract micropillars from mineralised turkey leg tendon and cold-water coral material [26,27]. The laser machining workstation was based on a Carbide laser (Light Conversion, Lithuania) operating at a wavelength of $\lambda$ = 1028 nm. Ultrashort pulses were delivered at a repetition frequency of 2 kHz with a full width half maximum pulse length of $\tau$ = 6 ps. An average power of 22 mW was used, corresponding to a pulse energy of 11 µJ. The beam waist radius of $\omega$ = 10 µm was focused on the surface of the sample. A galvanometer scan head was used to scan an inward Archimedean spiral pattern at a speed of 4.4 mm/s. The combination of the pulse repetition frequency and scan speed resulted in a beam overlap of 89%. The spiral pattern was repeated three times to cumulatively ablate material to increase the pillar aspect ratio. We measured the geometry of the micropillars using scanning electron microscopy (SEM) images acquired at two tilt angles. The micropillars exhibited average surface and base diameters of 32.2 ± 1.0 μm and 93.3 ± 2.6 μm, respectively. The average height was 143.1 ± 8.0 μm and the average taper angle was 12.3 ± 0.9°.

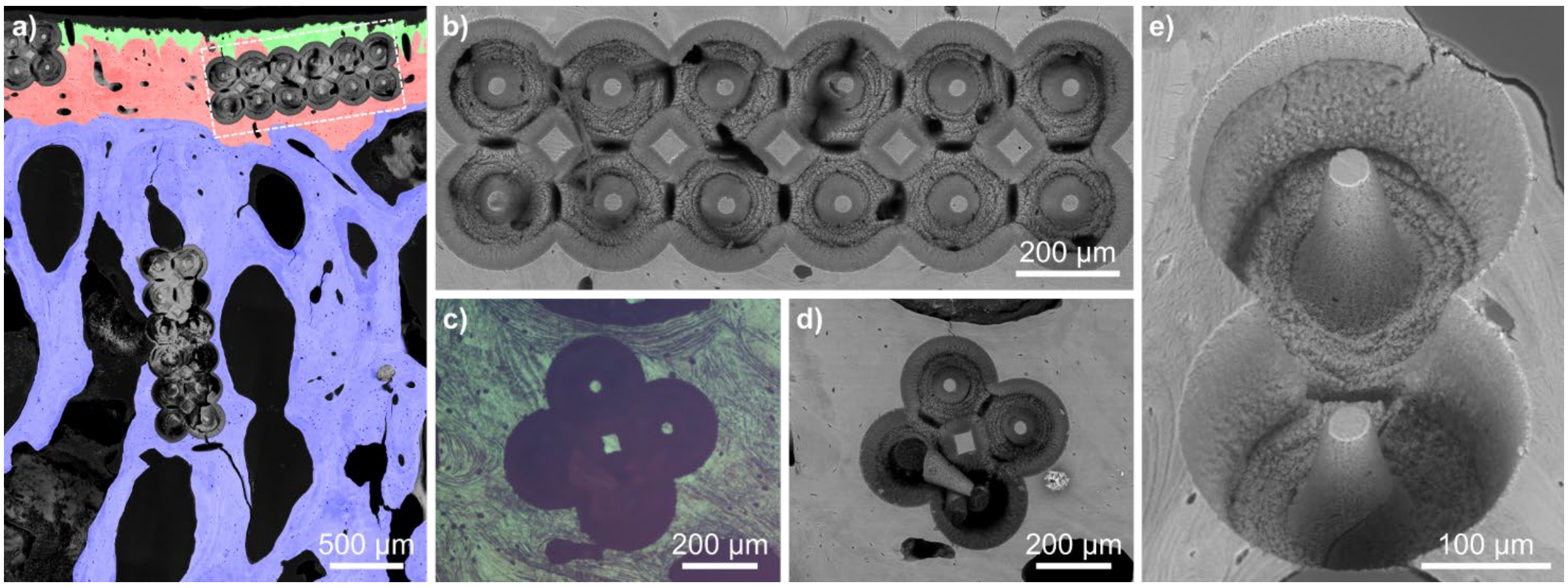

**Figure 1 – Subchondral bone micropillar arrays.** a) SEM image of subchondral bone unit with colour overlay, highlighting calcified cartilage (green), subchondral bone plate (red) and subchondral trabecular bone (blue). The white bounded box corresponds to- b) SEM image a typical 6x2 micropillar array. c-d) Light microscope and SEM image of an array with an intact toppled pillar, illustrating the achieved taper angle and aspect ratio. e) Angled projection SEM image illustrating the tapered profile of the pillar. The image also provides an example for an accepted and a rejected micropillar. The top pillar was deemed acceptable, while the bottom pillar was rejected due to visible porosity at the base of the pillar.

Arrays of micropillars were positioned in the subchondral bone plate and the subchondral trabecular bone of each embedded specimen (Figure 1a). Positioning of the arrays was done by referencing a stitched reflected light microscopy image of the sample surface. As the image only provided the in-plane spatial position of bone material at the surface, we were unable to assess the underlying through-plane composition of the material that composed each pillar when selecting the pillar location. For verification that each pillar consisted wholly of bone material without significant porosity or PMMA inclusion, we acquired low vacuum SEM images of each pillar array. Three investigators (S. McPhee, M. Peña Fernández, and U. Wolfram) then reviewed the SEM images and graded each pillar based on a binary categorisation of acceptance or rejection. Specific criteria for rejection included the presence of porosity within or under the pillar, the inclusion of PMMA in the pillar, or defects such as chips, cracks, or irregular geometry. Figure S1 (Supplementary materials) illustrates examples of rejected micropillars.

## 2.3. Quantitative backscattered electron imaging

To assess the degree of mineralisation within the embedded subchondral bone specimens, we evaluated the bone mineral density distribution (BMDD) by quantitative backscattered electron imaging (qBEI) after mechanical testing (Sections 2.5 and 2.6). We coated each sample by carbon rod evaporation after 72 hours under vacuum. BEI was performed on a digital scanning electron microscope (Quanta 650 FEG, Thermo Fisher, USA) equipped with 4 quadrant backscattered electron detector, and operated at 20 kV with a working distance of 15 mm. 16-bit BE images were acquired with a 2.36 mm horizontal field of view and a pixel size of 1.34 $\mu m^2$.

To accommodate quantification of the BE signal, we imaged a calibration standard composed of carbon and aluminium references (MAC Consultants Ltd., UK) before and after each bone specimen was imaged. The gain (contrast) and offset voltage (brightness) were first adjusted such that the grey-level of the carbon and aluminium regions exhibited a median grey-level value of approximately 15% and 85% of the 16-bit grey-level range [32]. Overlapping BE images (Figure 2a-b) covering the entire exposed bone region were acquired and stitched using 'MAPS' software (FEI, USA). Image post-processing was conducted in Python and ImageJ (NIH ImageJ 1.53 [33]), where we calibrated the grey-level values to a calcium weight percentage (Ca-Wt%) following the method of Roschger et al. [32]. For a pillar specific measure of the mineralisation, we evaluated the median Ca-Wt% of a 250 μm radius surrounding each pillar.

## 2.4. Raman spectroscopy

Raman spectra were collected for each pillar using a dispersive Raman microscope (inVia Reflex, Renishaw, UK) equipped with a 785 nm diode laser and 50× objective. The laser power at the sample surface was ~12 mW and spectra were acquired with a 30 s acquisition time with a single accumulation in the wavenumber range of 200–2000 $cm^{-1}$. Post-processing was done in Python v3.6, where we first baseline corrected each spectrum using an asymmetric least-squares fitting algorithm (pybaselines [34,35]) (Figure 2d). From the corrected Raman spectra, we evaluated a set of bone compositional properties. First, the mineral crystallinity, which is a relative measure of the crystallite size and is derived as the inverse of the full-width at half maximum (FWHM) of the $v_1PO_4^{3-}$ band (~960 $cm^{-1}$). We fitted the $v_1PO_4^{3-}$ band with a Lorentzian function with an additional linear baseline (Figure 2e). Second, we evaluated two intensity ratios within the Amide I band (~1660 $cm^{-1}$) for probing the organic phase.  We fitted the Amide I band with a triple Gaussian function with an additional linear baseline (Figure 2f). The second derivative of this fitted function was evaluated, and the indices of the local minima at 1640 $cm^{-1}$, 1670 $cm^{-1}$, and 1690 $cm^{-1}$ were used evaluate the intensity (functional value) at each position (Figure 2f). We then evaluated the intensity ratios of $I_{1670}/ I_{1640}$, which is indicative of collagen conformational change, and $I_{1670}/ I_{1690}$ which is indicative of matrix maturity [36,37].

## 2.5. Nanoindentation

Nanoindentation testing was performed on a Hysitron Triboindenter (Hysitron, USA) equipped with a diamond Berkovich probe and wet cell. The indentation protocol consisted of a load-controlled trapezoidal profile with an initial monotonic loading to 80 mN at a rate of 120 mN/min, succeeded by a holding period of 30 s and finished by unloading at a rate of 480 mN/min following pervious protocols [30]. The resultant indentation depth was approximately 2.5 µm. Each embedded specimen was rehydrated in Hank's Balanced Salt Solution (HBSS) for 90 minutes prior to testing and the wet cell ensured the specimen remained hydrated throughout testing. A total of 600 indents were programmed for both the ND and OA specimens, located in regions of transverse lamellae. We evaluated the reduced modulus ($E^*$) using Oliver and Pharr's method [38].

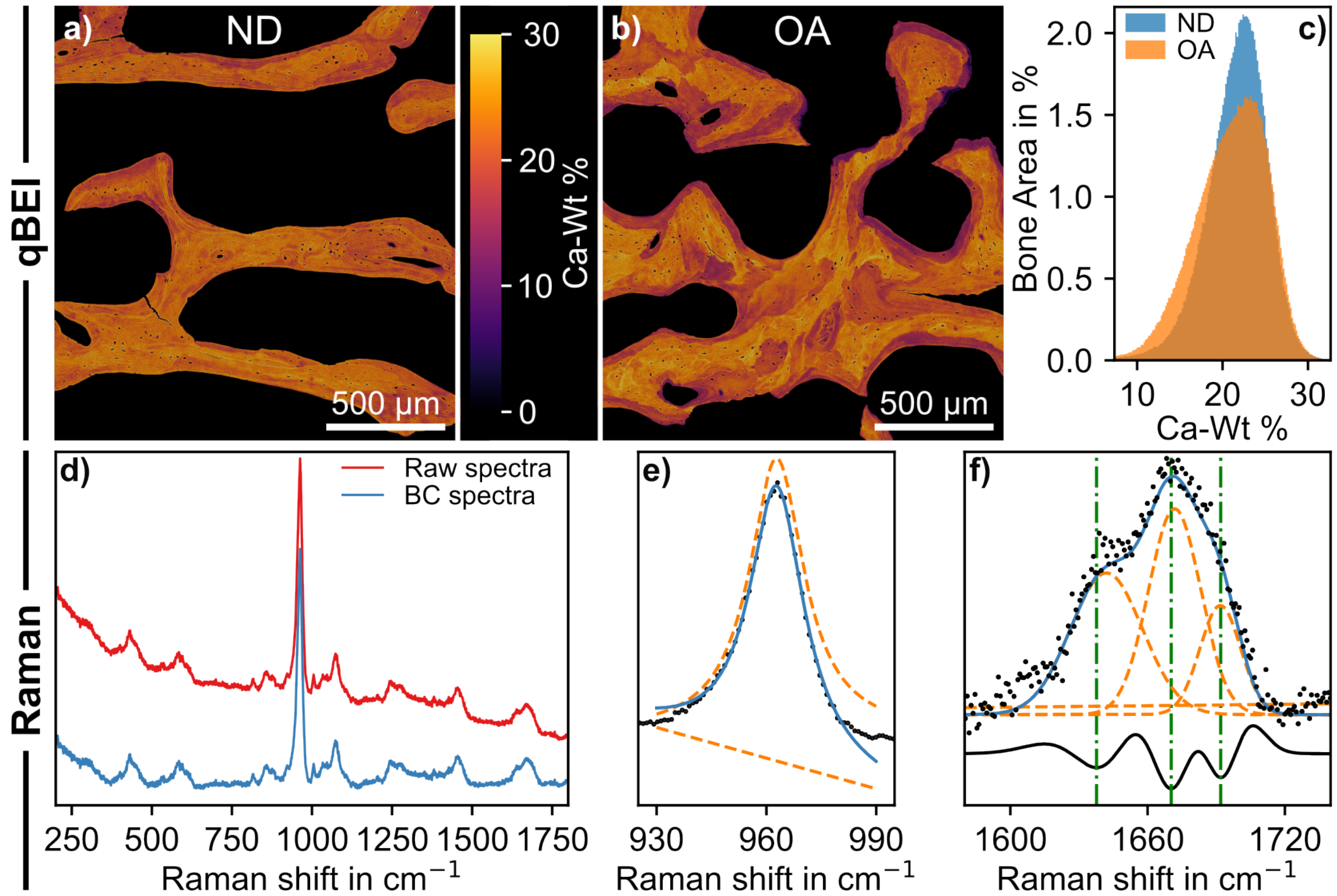

**Figure 2 – Compositional analysis by quantitative backscattered electron microscopy (qBEI) and Raman spectroscopy.** a) non-diseased (ND) and b) osteoarthritic (OA) trabecular bone sections respectively. The contrast and brightness are equal in both images such that the image intensities are comparable. c) histograms of the bone mineralisation density distribution BMDD as calcium weight percentage (Ca-Wt %) for the two ND and OA image sections. Note the mineralisation heterogeneity in the OA image, which is reflected by a broader BMDD. d) Example of a raw (red) and baseline corrected (blue) spectra acquired from a pillar. e) Lorentzian fitted $v_1PO_4^{3-}$ peak, used to determine the mineral crystallinity (1/FWHM). f) Amide 1 band fitted with three gaussian curves, used to determine the functional value of $I_{1670}/I_{1640}$ and $I_{1670}/I_{1640}$. In e and f, the blue curve indicates the composite function of the underlying individual fitting functions (dashed orange).

## 2.6. Micropillar compression

Each micropillar was compressed using a custom-made portable microindenter (Alemnis AG, Switzerland) fitted with an 88 µm diameter flat punch probe and custom-made liquid cell. Before testing, we rehydrated each specimen by submerging them in Hank's buffered saline solution (HBSS, Sigma Aldrich, USA) for a period of 90 minutes, similar to the rehydration in case of nanoindentation. Each pillar was then compressed uniaxially to a total displacement of 15 µm at a quasi-static displacement rate of 0.05 µm/s. After an initial displacement of 2 µm, partial unloading cycles were performed by retracting the probe 0.25 µm for each successive 1 µm displaced by the probe. A graphical visualisation of the loading protocol can be found in Figure S2 in the Supplementary materials. Throughout each test, displacement, force, and elapsed time were recorded at a sampling frequency of 30 Hz.

Unlike in straight micropillars [19,24] The tapered profile of the pillars prevented direct conversion of the force-displacement data to a uniaxial stress-strain curve. To determine the micromechanical elastic- and yield properties, we developed a numerical back-calculation routine by conducting *in silico* micropillar compression tests. To facilitate this, we first extracted relevant characteristic parameters from the force-displacement data (Figure 3a) as targets for the numerical investigation. First, the displacement was compliance corrected to account for frame compliance [39]. Each curve exhibited a toe-region, which is an inherent initial progressive stiffness increase caused by an initial incomplete contact between the pillar and probe mostly due to surface roughness and preparation artefacts [40]. To account for this, we follow Indermaur et al. [41] by first evaluating the local stiffness (kernel size = 5) and extrapolating the maximum loading stiffness ($K_{load}$) to find the initial displacement at which the force is zero. The displacement was then corrected to account for this initial loading nonlinearity (Figure 3a).

Following the initial displacement correction, we evaluated a pseudo-yield force ($F_y$) with a 0.2% offset criterion (Figure 3a). To evaluate the apparent stiffness $\left(K_{app}\right)$ of each pillar, we fitted the last unloading segment prior to yield [26] with a third-order polynomial. $K_{app}$ was then the gradient of the tangent evaluated at the initial unloading displacement (Figure 3a-b). Finally, the gradient of the force-displacement curve post-yield ($K_{py}$) was evaluated (Figure 3a).

## 2.7. *In silico* micropillar compression

With $K_{load}$, $F_y$, $K_{app}$, and $K_{py}$ evaluated from the experimental force-displacement data of each compressed micropillar, we set out a numerical investigation to back-calculate the respective Young's modulus ($E_0$) and compressive yield stress ($\sigma_0^-$). To do so, we defined a finite element model to perform *in silico* micropillar compression. To back-calculate the micromechanical parameters of the constituent material within each micropillar, we minimised the error between the experimental- and numerical force-displacement curves.

We generated a mesh with dimensions consistent with the average micropillar, which sat atop a 150 µm diameter hemisphere to accommodate pillar sink-in [27,42]. The mesh was equipped with an elasto-viscoplastic material model, proposed by Schwiedrzik et al. [43], as a user defined material subroutine (UMAT) in Abaqus (v6.16, Dassault Systèmes). The constitutive equations are detailed in Supplementary materials S1. The elastic domain exhibits isotropic stiffness and is bounded by an asymmetric quadric yield surface [44], which takes the form of a cone with a rounded tip that is aligned with the hydrostatic stress axis in normal stress space (see Supplementary materials Figure S3). The yield criterion accounts for the asymmetry in tensile and compressive strength, which is evident in bone tissue at the macro- [45,46] and microscale [47]. We used a constant ratio of $\sigma_0^+ = \frac{2}{3}\sigma_0^-$ [45–47], with the magnitude $\sigma_0^-$ as input. We implemented post-yield linear hardening, where in addition to evaluating $E_0$ and $\sigma_0^-$, we also evaluated the post-yield gradient, $m_{py}$. The influence of varying the input parameters ($E_0$, $\sigma_0^-$ and $m_{py}$) on the uniaxial stress-strain behaviour is illustrated in Supplementary materials Figure S4.

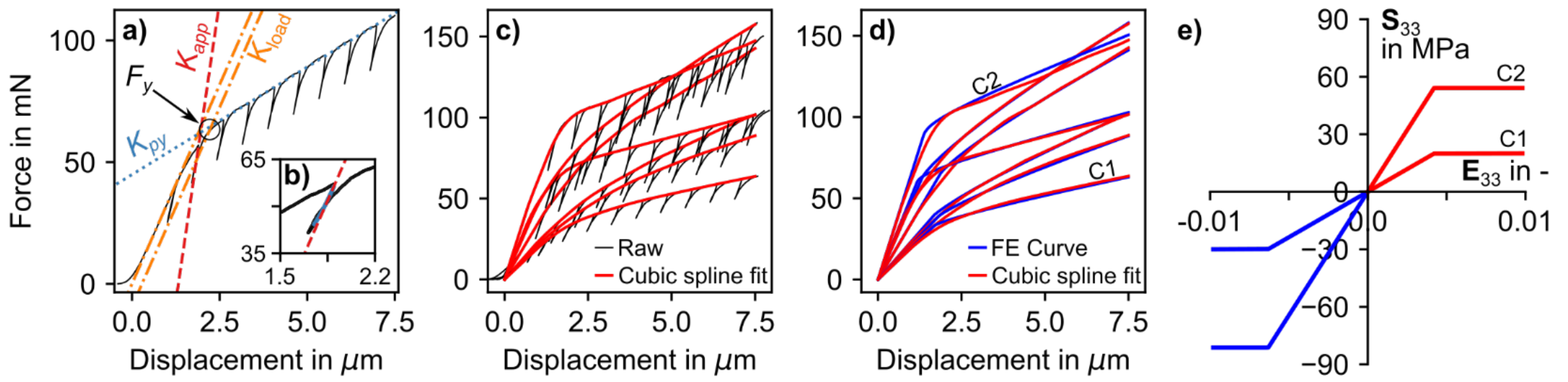

**Figure 3 – Micropillar compression force-displacement results.** a) An exemplary force-displacement curve from which the following variables were evaluated: apparent stiffness ($K_{app}$, red dashed line); loading stiffness ($K_{load}$, orange dot-dashed line); post-yield gradient ($K_{py}$, blue dot line); and yield force ($F_y$, circled). Note the initial loading nonlinearity has been corrected such that the plot starts at a negative displacement. b) Subplot of the unloading curve prior to yield, with the fitted third-order polynomial (blue solid curve) and $K_{app}$ (red dashed line). c) Seven exemplary curves which all exhibit close to the median compressive yield strain, but with a range of stiffnesses and strengths. The black curve here is the raw data with unloading segments present, while the overlying red curves are cubic splines fitted to the initial unloading points. d) The same curves in a) with the FE-derived force-displacement curves in blue present. e) Uniaxial stress-strain behaviour of the material definition that comprises the pillar for which the force-displacement curves are composed in d), where the labelled curves C1 and C2 are correspondent. Notice the asymmetry in tensile (red) and compressive (blue) yield strength.

Given that the diamond flat punch has a stiffness of ~1141 GPa, deformation of the modelled micropillar was accommodated through contact with a rigid body with displacement control boundary conditions assigned. During experimental micropillar compression, some localised plastic deformation occured at the interface between the micropillar and probe. This was evidenced by the loading stiffness being lower than the unloading stiffness, the latter of which is considered a response of entirely elastic recovery. To correct for this reduced loading stiffness, we implemented a user-defined contact constraint (UINTER) in Abaqus. During loading, contact was modelled using a contact stiffness, $K_{cont}$, allowing the rigid body to overclose (penetrate) the pillar surface. Upon unloading, the contact stiffness was increased to constrain the overclosure during probe retraction. This softened contact during loading reduces loading stiffness. It must be noted that this method simply corrects the numerical force-displacement curve and does not influence the mechanical response of the pillar. This correction is necessary to apply a comparative yield criterion to the numerical force-displacement curve.

For each micropillar, $E_0$, $K_{cont}$, $\sigma_0^-$ and $m_{py}$ were derived iteratively using a secant method to minimise the relative error, $|1-(Y_{FE}/Y_{EXP})|$, between the experimental and numerical force-displacement curve. Here $Y$ is one of the target variables $E_0$, $K_{cont}$, $\sigma_0^-$ or $m_{py}$. A detailed description of the method is found in Supplementary materials S2. To assess the accuracy of the numerical back-calculation routine, we quantified the mean absolute percentage error (MAPE), the root mean squared error (RMSE) and maximum absolute error (MaxAE) between the experimental and converged numerical force-displacement curves for each micropillar.

## 2.8. Data analysis

Statistical analyses were conducted using Python with Scipy and Statsmodels [48] , and R with lme4, lmerTest and emmeans. To compare the micropillar compression micromechanical and compositional variables between location and disease state, while accounting for repeated measurements across samples, we employ a mixed-effect model. In this model, disease state (ND and OA) and location (BP and TB) are treated as fixed effects, and samples treated as a random effect as:

$$v = s * l + (\ 1 \mid \Sigma) \tag{1}$$

with variable $v$, state s, location l, and sample $\Sigma$.

For the nanoindentation variable, location is omitted to reflect the absence of measurements in the subchondral bone plate. Differences in the fixed effect groups were tested via ANOVA type analysis. Estimated marginal means (EEMeans) were estimated and pairwise comparisons were performed as a post-hoc test [49]

To establish whether a relationship existed between the micropillar compression derived mechanical properties and the Raman and qBEI compositional measurements. Following Mirzaali et al [30], we used a power law function with the qBEI and Raman variables as explanatory variables as:

$$X = X_0 \left(\frac{\text{Ca} - \text{Wt\%}}{\overline{\text{Ca} - \text{Wt\%}}}\right)^a \left(\frac{FWHM_{\sim 960}^{-1}}{\overline{FWHM_{\sim 960}^{-1}}}\right)^b \left(\frac{MMR}{\overline{MMR}}\right)^c \left(\frac{\sim I_{1670}/\sim I_{1640}}{\overline{\sim I_{1670}/\sim I_{1640}}}\right)^d \left(\frac{\sim I_{1670}/\sim I_{1690}}{\overline{\sim I_{1670}/\sim I_{1690}}}\right)^f \tag{2}$$

Here $X$ represents one of the FE-derived micromechanical parameters ($E_0$, $\sigma_0^-$ and $\varepsilon_0^-$) and the Raman and qBEI measurements were normalised to the population averages (denoted by an overbar) such that $X_0$ represents a fictitious material with average composition as measured by qBEI and Raman. $X_0$ and the exponents $a$, $b$, $c$, $d$ and $f$ were then fitting parameters.

We utilised a log-space conversion of Equation 2 and performed ordinary least squares regression, where we sequentially removed non-significant variables, each time removing the variable with the highest p-value until only significant variables remained. For all tests, we assumed a significance level of $\alpha = 0.05$.

# 3. Results

## 3.1. Micromechanical testing

A total of 529 (ND = 266, OA = 263) indentations were successfully performed in subchondral trabeculae under hydrated conditions (Figure 4). Nanoindentation revealed a significant dependence ($p = 0.0311$) of reduced modulus on disease state in trabecular bone.

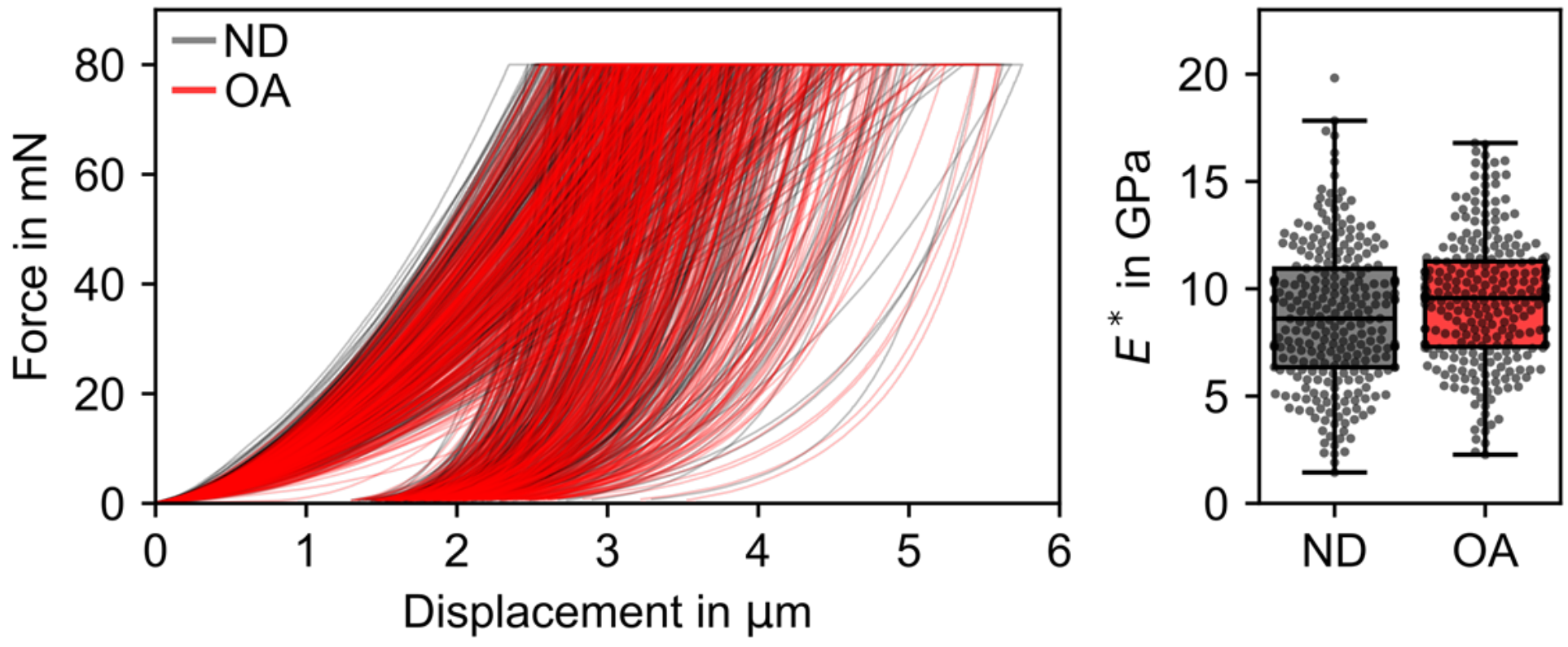

**Figure 4 – Nanoindentation results pooled for the entire population.** Force displacement data showing loading, hold period and unloading phases for all nanoindentations for ND (grey) and OA (red) subchondral trabecular bone. The box plot illustrates the resultant reduced modulus $E^*$.

A total of 336 micropillars were programmed for laser ablation extraction from which 227 were successfully extracted. From these, 109 pillars were deemed acceptable by all three investigators and only these pillars were subject to subsequent analyses. Figure 3b shows a range of exemplary force-displacement curves. Beyond yield, the pillars predominantly exhibited a linear hardening-like behaviour (Figure 3b-c).

The back-calculated micromechanical properties $E_0$, $\sigma_0^-$ and $\varepsilon_0^-$ are shown in Figure 5. Micropillar compression revealed no significant dependence of $E_0$ on the micropillar location or disease state (Table 1). The significant interaction between disease state and location for $\sigma_0^-$ was a consequence of an increase in trabecular bone $\sigma_0^-$ in OA ($p$ = 0.031, EEMeans pairwise comparisons) that is absent in ND ($p$ = 0.98, EEMeans pairwise comparisons). $\varepsilon_0^-$ was significantly higher in OA ($p$ = 0.041, EEMeans pairwise comparisons), but not dependent on location ($p$ = 0.996, EEMeans pairwise comparisons).

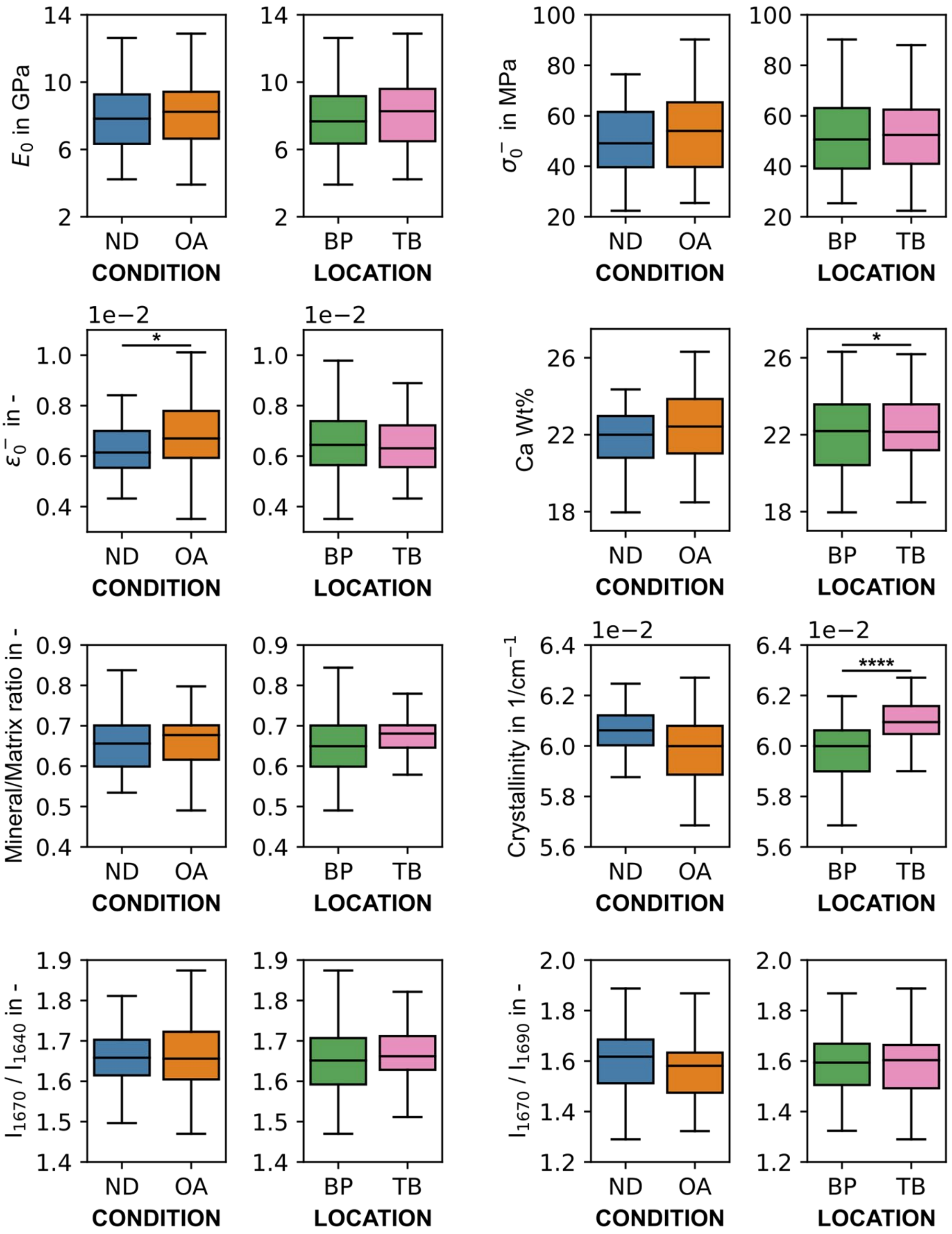

**Figure 5 – Micropillar compression and compositional results.** Box plots of the micromechanical and compositional properties of subchondral bone in relation to disease state (non-disease, ND and osteo-arthritis, OA) and location (trabecular, TB or bone plate, BP). See Table 1 for mixed effect model results. Significance level notations: '*' - $p$ < 0.05, '**' - $p$ < 0.01, '***' - $p$ < 0.001, **** - $p$ < 0.0001.

**Table 1 – Micromechanical and compositional measurements of subchondral bone** by disease state (non-disease, ND and osteoarthritis, OA) and location (trabecular, TB or bone plate, BP). Values are reported as mean ± standard deviation. The reported $p$-values relate to the one-way ANOVA test with the mixed effect model where disease state (state), location, and their interaction (Interact.) are reported.

| Method | Value | Disease State | | Location | | p-values | | |
|---|---|---|---|---|---|---|---|---|
| | | ND | OA | BP | TB | State | Location | Interact. |
| **Nanoindentation** | No. measurements | n = 263 | n = 248 | - | - | | | |
| | $E^*$ in GPa | 8.71 ± 3.26 | 9.47 ± 2.95 | - | - | **0.031** | - | - |
| **Micropillar Compression** | No. measurements | n = 60 | n = 49 | n = 62 | n = 47 | | | |
| | $E_0$ in GPa | 7.97 ± 2.16 | 8.17 ± 2.29 | 8.01 ± 2.26 | 8.14 ± 2.17 | 0.450 | 0.499 | 0.109 |
| | $\sigma_0^-$ in MPa | 50.13 ± 14.41 | 54.6 ± 16.41 | 51.94 ±15.61 | 52.46 ± 15.36 | 0.309 | 0.161 | **0.031** |
| | $\varepsilon_0^-$ in ($10^{-2}$) | 0.633 ± 0.114 | 0.679 ± 0.140 | 0.659 ± 0.134 | 0.649 ± 0.121 | **0.041** | 0.996 | 0.117 |
| **Raman** | No. measurements | n = 60 | n = 49 | n = 62 | n = 47 | | | |
| | $MMR$ in - | 0.65 ± 0.07 | 0.67 ± 0.08 | 0.65 ± 0.08 | 0.67 ± 0.06 | 0.107 | 0.315 | 0.240 |
| | $FWHM_{\sim960}^{-1}$ in - | 0.0606 ± 0.0009 | 0.0599 ± 0.0013 | 0.0598 ± 0.0011 | 0.0609 ± 0.0010 | 0.082 | **<0.0001** | **0.033** |
| | $I_{1670}/I_{1640}$ in - | 1.66 ± 0.07 | 1.66 ± 0.091 | 1.65 ± 0.09 | 1.66 ± 0.07 | 0.914 | 0.930 | 0.189 |
| | $I_{1670}/I_{1690}$ in - | 1.60 ± 0.14 | 1.57 ± 0.12 | 1.58 ± 0.12 | 1.59 ± 0.14 | 0.946 | 0.145 | 0.473 |
| **qBEI** | No. measurements | n = 60 | n = 49 | n = 62 | n = 47 | | | |
| | Ca-Wt% in % | 21.72 ± 1.66 | 22.49 ± 1.94 | 21.95 ± 1.94 | 22.22 ± 1.68 | 0.502 | **0.032** | 0.405 |

The median MAPE, RMSE and MaxAE between the numerical and experimental curves were 1.62%, 1.63 mN and 4.40 mN respectively. Similarly, the 95th percentiles were 5.44%, 5.91 mN and 10.8 mN. Eight micropillars exhibited a MaxAE above 10 mN. These higher errors stem from a post-yield softening behaviour that may be due to internal flaws such as cellular porosity. These pillars have not been excluded from the evaluation since the larger error is due to the low agreement with the hardening behaviour in the model. The elastic modulus and yield strength for these pillars where therefore considered usable.

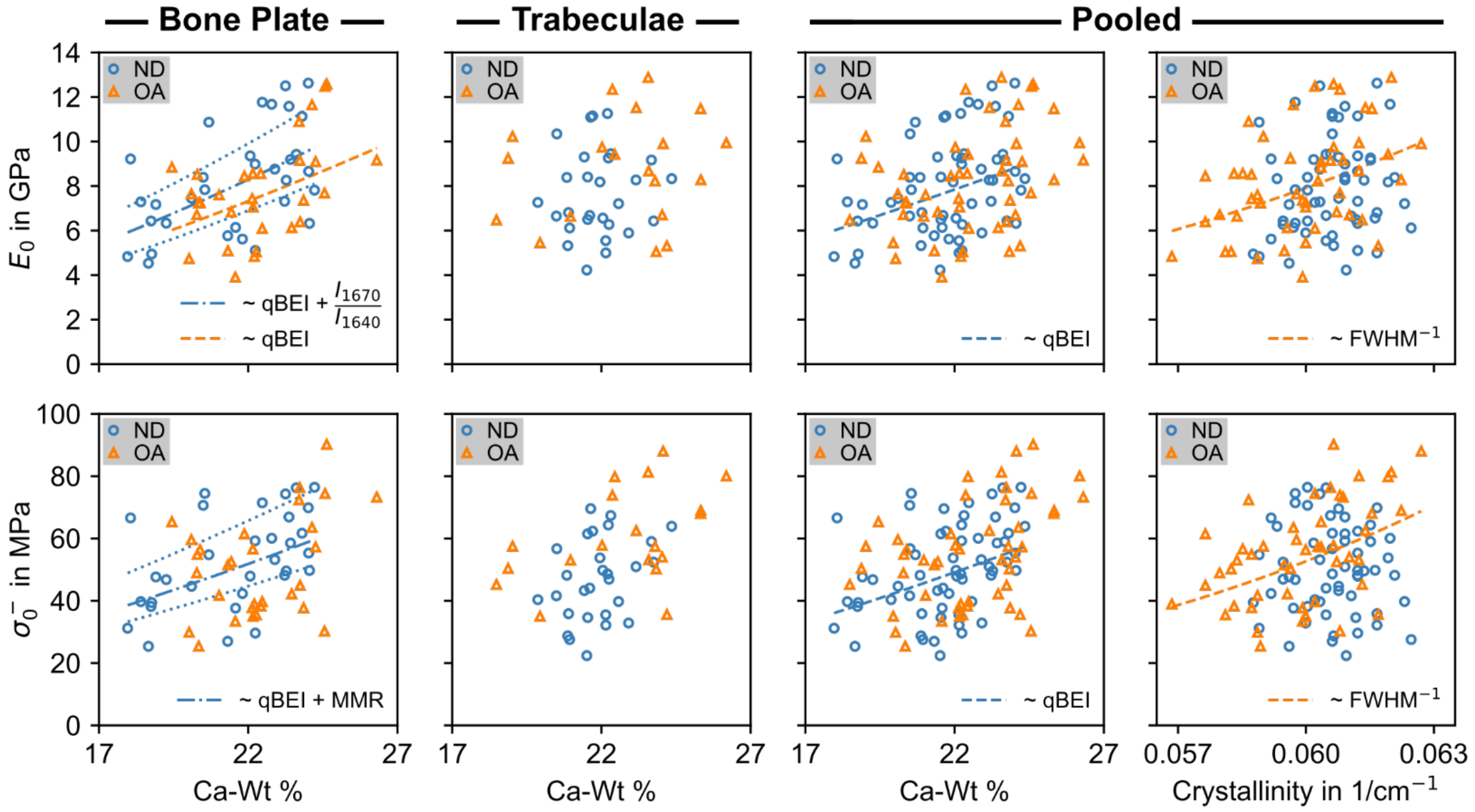

**Figure 6 – Elastic modulus and compressive yield strength as a function of compositional variables.** The orange triangles and blue circles represent the OA and ND measurements respectively. A dashed line corresponds to a single variable power law function (Equation 2, Table 2). The dotted and dot-dashed lines correspond to a two variable power law function (Equation 2, Table 2), where the first independent variable is continuous while the second variable is fixed at its median (dot-dash) and minimum and maximum (dot).

## 3.2. Compositional analysis

Pillar specific Raman spectroscopic and qBEI results are shown in Figure 5, and Table 1. Raman spectroscopy revealed that mineral crystallinity is location dependent ($p < 0.0001$). The significant interaction between disease state and location was a consequence of a significant reduction between ND and OA in the subchondral bone plate ($p = 0.012$, EEMeans pairwise comparisons) and not in the trabeculae ($p = 0.679$, EEMeans pairwise comparisons). The Amide I intensity ratios showed no location or disease state dependency ($p > 0.05$).

The results of the power law correlation analysis are outlined in Table 2. The analysis revealed that in the bone plate, the qBEI median Ca-Wt% was the dominant variable on which the micromechanical properties are dependent (Figure 6, Table 2). When the micropillar location were pooled, there existed a dependence of $E_0$ and $\sigma_0^-$ of the OA micropillars on the mineral crystallinity.

Table 2 – Significant ($p$<0.05) fitting parameters of the power law regression analysis (Equation 2). $X_0$ corresponds to a material with average compositional properties. The exponents $a$, $b$, $c$, $d$ and $f$ correspond to $\mathrm{Ca-Wt\%}$, $FWHM_{\sim 960}^{-1}$, $MMR$, $\sim I_{1670}/{\sim}I_{1640}$, ${\sim}I_{1670}/{\sim}I_{1690}$ variables.

| $X$ | Location | State | $X_0$ | $a$ | $b$ | $c$ | $d$ | $f$ | Adj. $R^2$ |
|---|---|---|---|---|---|---|---|---|---|
| $E_0$ in GPa | BP | ND | 8.41 | 1.66 | - | - | 1.90 | - | 0.331 |
| | | OA | 7.36 | 1.58 | - | - | - | - | 0.153 |
| | TB | ND | 7.52 | - | - | - | - | - | - |
| | | OA | 8.12 | - | - | - | - | - | - |
| | Pooled | ND | 7.88 | 1.32 | - | | | - | 0.127 |
| | | OA | 8.12 | - | 5.21 | | | - | 0.146 |
| $\sigma_0^-$ in MPa | BP | ND | 55.00 | 1.46 | - | 0.96 | - | - | 0.299 |
| | | OA | 47.76 | - | - | - | - | - | - |
| | TB | ND | 45.91 | - | - | - | - | - | - |
| | | OA | 57.25 | - | - | - | - | - | - |
| | Pooled | ND | 49.37 | 1.53 | - | - | | - | 0.137 |
| | | OA | 54.26 | - | 6.05 | - | | - | 0.174 |
| $\varepsilon_0^-$ in - | BP | ND | 0.658 | - | - | - | - | - | - |
| | | OA | 0.652 | - | - | - | - | - | - |
| | TB | ND | 0.614 | 1.70 | - | - | - | - | 0.209 |
| | | OA | 0.681 | - | - | - | - | - | - |
| | Pooled | ND | 0.649 | 0.61 | -5.70 | - | - | - | 0.179 |
| | | OA | 0.662 | - | - | - | - | - | - |

# 4. Discussion

In this study, we used a combination of nanoindentation, micropillar compression testing, and compositional analyses to assess microscale elastic properties and yield strength of rehydrated subchondral bone in non-diseased and late-stage ankle osteoarthritis samples. By combining micromechanical and tissue compositional analyses we were able to investigate how the mechanical properties are related and how these properties are affected in subchondral bone by osteoarthritis (Figure 6, Table 2).

## 4.1. Micromechanical testing

Nanoindentation revealed reduced moduli (Figure 4) for OA trabecular bone (9.45 ± 2.95 GPa) and ND bone (8.71 ± 3.26 GPa) in these male distal tibial samples. These values compare with Wolfram et al. [50], who conducted indentations on rehydrated lamellae in a transverse orientation in human vertebral trabecular bone. We observed a 0.81 GPa (95% CI [0.05, 1.58]) increase in reduced modulus in late-stage OA bone compared to ND bone. This increase is consistent with Renault et al. [16] who observed a correlation between rehydrated subchondral trabecular bone indentation modulus and bone volume fraction. This is indicative of disease progression [51]. Similarly, Zuo et al. [18] observed an increased in indentation modulus, albeit between early (KL-score = 1) and late-stage (KL-score = 4) OA trabecular lamellae. Interestingly, Renault et al. [16] attribute this positive association to the maturity of the bone. The increase in thickness via remodelling is a surface phenomenon where new lamellar packets are deposited on the periphery, while the centre of each trabecula is comprised of more mature bone. We observed this in the qBEI of OA specimens, where hypomineralised packets of lamellae were located on the outer peripheries of individual trabeculae, consistent with newly formed bone (Figure 2). Indentations were always positioned centrally to any incident trabeculae and as such, we unlikely probed these regions of hypomineralised bone. We believe the results for both the nanoindentation and micropillar compression must be viewed in light of this positioning and we revisit this concept later.

Assuming a Poisson's ratio of $\nu_0 = 0.3$ for bone and $\nu_d = 0{,}07$ for diamond, the reduced moduli we measured using nanoindentation (Figure 4 and 5) would translate into a Young's modulus of $E_0 = 8{,}67$ GPa and $E_0 = 7.99$ GPa for OA trabecular bone and ND bone, respectively. Therefore, the micropillar compression derived $E_0$ values were comparable to the nanoindentation reduced modulus values, albeit without significant difference related to the disease state of the tissue (Figure 5). Our indentation results are ~20% - 50% lower than other studies on hydrated and non-hydrated lamellae in axially loaded compact bone [25,52,53]. This reduction may be explained by lower transverse stiffness compared to axial stiffness as observed by Wolfram et al. [50]. Furthermore, Hengsberger et al. [52] demonstrated that indentation modulus decreases with increasing indentation depth. In Wolfram et al. [50] and Renault et al. [16], the loading protocols were comparable in magnitude to ours, where an indentation depth of ~2.5 μm averages across approximately three lamellae [50]. In Hengsberger et al. [52] and Guidoni et al. [53], however, the maximum indentation load was ~6% of ours. Schwiedrzik et al. [25] tested dry tissue to a maximum depth of 1 μm, which is roughly a third compared to our indentations. Additionally, Schwiedrzik et al. [25] and Guidoni et al. [53] both tested in cortical bone, which was shown by Zysset et al. [54] to have a cortical-to-trabecular elastic modulus ratio of ~1.76. The stiffness observed here fits to resources that reported results on transversally oriented indentations [16,50]. The reduced stiffness in comparison to other studies can be attributed to different material orientation, greater indentation depth, and difference in indentation locale. Furthermore, we believe that obtaining consistent results by micropillar compression and nanoindentation, which are mutually complementary techniques, underscores the validity of our results.

We showed that not only stiffness, but also tissue yield stress in ankle OA subchondral trabecular bone is increased in OA compared to non-disease (Table 1). To the best of our knowledge, micromechanical compressive yield strain and stress of subchondral trabecular bone and the bone plate in healthy and OA human ankle tissue have not yet been reported. As such, we cannot directly compare the measured yield properties with previous studies. Our measured yield stress values (50.13 ± 14.41 MPa in ND and 54.6 ± 16.41) were significantly lower than values reported by Kochetkova et al. [22], who tested similar sized micropillars in femoral neck cortical bone (257.9 MPa). However, Kochetkova et al. [22] tested micropillars in dry conditions and it has previously been shown that rehydration has a significant effect on microscale yield properties of bone with up to three fold decreases observed [19,20,55]. Furthermore, yield properties of subchondral bone may differ from those of cortical bone. Indeed, others have reported lower tissue strength values for trabecular bone tissue in comparison to cortical bone [56]. This may be explained by differences in lamellar architecture and mineralisation. Tertuliano and Greer [57] observed compressive yield stress values > 300 MPa in 3 µm diameter micropillars of trabecular bone tissue from human femoral condyle. However, not only did the authors perform compression experiments in dry conditions but also illustrated the drastic effect of micropillar size on yield stress values, which extends earlier findings for cortical bone [25]. At larger length scales, Frank et al [58] reported tensile yield stress values of ~ 30 MPa in individual trabeculae from the human femoral neck, which aligns with our results when considering previously reported tension-compression strength ratios of 0.4 [56] and 2/3 [59]. Finally, our compressive yield strain compares with [19,55]. Yield strain in bone tissue shows a small variation across different tissues [56,60,61]. For a thought experiment, we assume yield strain to be constant across trabecular and cortical tissue at the microscale, say $\varepsilon_0 \approx 0.7\%$ (Figure 5 and [19,55,61]). The yield point marks the end of the elastic regime so that we could estimate yield stress as $\sigma^y = E_0\, \varepsilon_0$. If we furthermore assume transverse stiffness of cortical bone to be ~20 GPa [19] and use our transverse stiffness of ~8.33 GPa, then we obtain 140.0 MPa and 58.3 MPa for cortical and trabecular tissue which fits to our compressive results for rehydrated tissue and those reported elsewhere [19,55]. Therefore, the difference in stiffness between cortical and trabecular tissue serves as an explanation for the differences in yield strength observed here for rehydrated subchondral trabecular bone and yield strength observed for rehydrated cortical tissue. We conclude that our measured yield strengths are sound.

## 4.2. Influence of porosity, orientation, and geometry on mechanical properties

The results from both indentation and micropillar compression share a large coefficient of variation, CoV, of ~30%. Bone tissue itself will possess some degree of inherent variation [62], which was shown to be exacerbated by hydration [50]. The relatively large indentation depth (~3 μm) and pillar size likely compounded this variation. Kochetkova et al. [28] similarly observed a high variation when conducting micropillar compression of cortical bone tissue with laser ablated micropillars in the size order of our pillars. They postulate that the size order of their pillar allows for a high degree of sub-microstructural inclusions within the pillars themselves. This rationale is plausible, and we have further investigated the potential influence of such sub-microstructural inclusions.

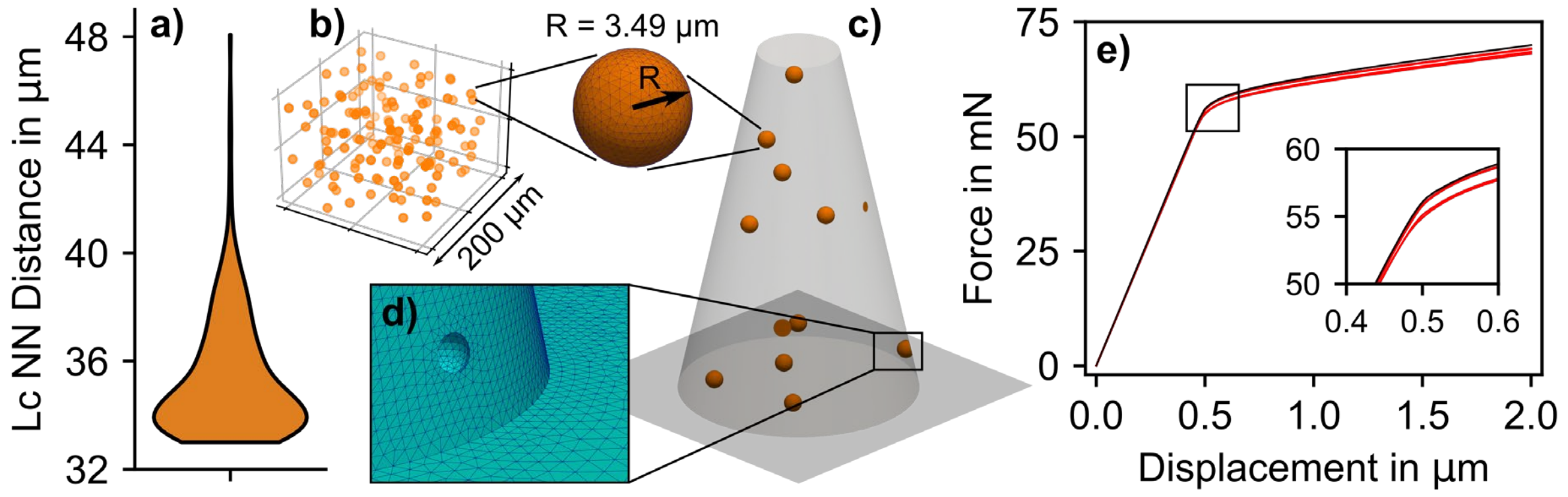

**Figure 7 – Influence of lacunar porosity on the micropillar response.** a-b) Distribution and corresponding violin plot displaying the nearest neighbour (NN) distance between any lacuna (Lc). c) Random positioning of lacunae within the micropillar. d) Example of the porosity imposition on the micropillar as a mesh feature where in this shown instance the lacuna intersects the exterior of the micropillar. e) Force-displacement data from *in silico* micropillar compression where the black curve represents a poreless micropillar and the five red curves are simulations with five random distributions of osteocyte lacuna.

Firstly, we considered if the inclusion of osteocyte lacunae would compromise the integrity of the micropillars (Figure 7). Using lacunar morphometric parameters reported by Goff et al. [63], who report a median lacuna volume of 178 μm$^3$ and degree of anisotropy of ~3.4. We used Poisson-disk sampling (scipy.stats.qmc.PoissonDisk [64]) to produce a point distribution with a number density of approximately 16.6 (1000/mm$^3$) to represent the spatial distribution of lacunae in trabecular bone (Figure 7a-b). Despite the high degree of anisotropy, for simplicity and because we loaded uniaxially, each point was first assigned a spherical volume of radius 3.49 μm [63] (Figure 7c). A Boolean difference operation on the computational domain was then performed such that any sphere that intersected the pillar region was removed and the porosity was explicitly imposed as mesh feature (Figure 7c-d). We performed five *in silico* micropillar compression tests to evaluate pore influence (Figure 7e), of which, the maximum reduction in stiffness and yield force was 0.99% and 2.83% respectively. With such marginal reductions here, we believe that inclusions such as lacunae have no consequence with respect to the variation of our micromechanical testing results. The minimal reduction in stiffness and strength did not warrant further investigating anisotropy of lacuna porosity as an ellipsoidal inclusion.

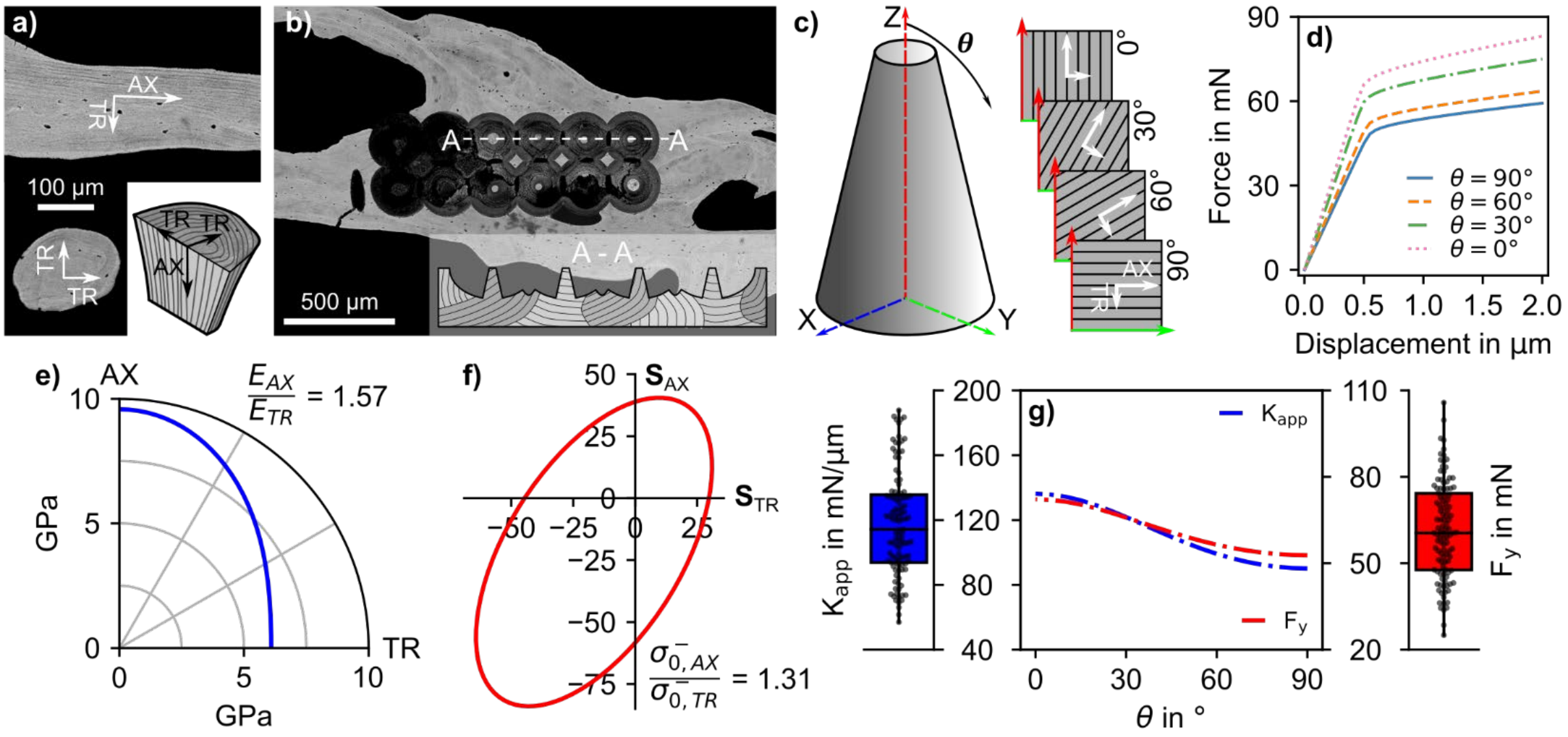

**Figure 8 – Influence of lamella anisotropy.** a) Areas of bone where orientation of lamellae may be inferred as transverse, TR, (top) or axial, AX, (bottom). b) Example of trabeculae juncture where pillars are located. The transect A-A corresponds to the lower right illustration of how the lamellae could transect the micropillars. c) Schematic of a pillar with global coordinate system X, Y, Z. The material orientation is then rotated through $\theta$ relative to the pillar axis (Z-axis). d) Resultant force-displacement curves for respective material orientations. e) Polar plot of elastic modulus ($E$) in the axial-transverse plane. f) Quadric yield surface in normal stress space in the axial-transverse plane in MPa. g) Apparent stiffness ($K_{app}$) and yield force ($F_y$) and as a function of θ. The box plots correspond to the experimental $K_{app}$ and $F_y$.

Lamellar bone exhibits significant anisotropy with regards to its elastic stiffness and strength [19,24,25,47,50], owing to the orientation of the underlying mineralised collagen fibres in the lamellae [65,66]. The size of our micropillars made confidently positioning micropillars in regions with a consistent lamellar orientation challenging. We positioned pillars close to or in junctions where trabeculae converge to ensure sufficient support materials and enough material to create the pillars (Figure 8b), and we had no means to assess the orientation of the lamellae within each pillar. To investigate the potential impact of an elastic and strength anisotropy, we substituted our original isotropic material definition with a transversely isotropic case, parameterised with axial (AX) to transverse (TR) elastic and strength ratios reported by Schwiedrzik et al. [19] ($E_{AX}/E_{TR}$ = 1.57, $\sigma_{0,AX}^-/\sigma_{0,TR}^-$ = 1.31). The formulation and parameter selection for the transverse isotropic stiffness and strength tensors can be found in Supplementary materials S3. Following principles outlined in [24,47,67], we rotated the axial material orientation with respect to the pillar axis and assessed the reduction in the resultant force-displacement curves as well as their respective stiffness and yield force (Figure 8c-d). With a fixed material parameterisation, varying only the alignment of material orientation with regards to the pillar axis from axial to transverse translated to a pillar stiffness and yield force reduction of 33.8% (1 - $K_{app,AX}/K_{app,TR}$) and 26.9% (1 - $F_{y,AX}/F_{y,TR}$). The variation in pillar stiffness and yield force, observed across axial and transverse orientations (AX: $K_{app}$ = 136.2 mN/µm, $F_y$ = 72.2 mN; TR: $K_{app}$ = 90.1 mN/µm, $F_y$ = 52.8 mN) captures the interquartile range of measured pillar stiffnesses (IQR 93.7 mN/µm - 135.6 mN/µm) and accounts for 73% of the interquartile range in experimental yield forces (IQR 47.7 mN - 74.3 mN), calculated as ($F_{y,AX}$ - $F_{y,TR}$)/IQR (Figure 8g). We may infer then that a large proportion of the experimental variability stems from lamella anisotropy. Addressing this in further works would require either individual positioning of micropillars in the areas where the lamella orientation can be confidently identified (Figure 8a, see e.g. Wolfram et al. [50]), or identifying the orientation of the lamellae in each pillar, for example by polarised Raman spectroscopy following methods proposed by Kochetkova et al. [24]. The drawback of the former would be the increase in fabrication time, potentially hindering the viability of laser ablation as high throughput capable extraction technique.

In addition to inclusion of lacunae and lamella anisotropy, we investigated the potential effect of geometric variability. The variability of the micropillars produced by laser ablation will include variations in height, taper angle, base and top diameters, and eccentricity. The micropillar geometry was previously fixed to represent the average micropillar dimensions, ensuring that only the micromechanical properties of the constituent material were investigated. For a qualitative evaluation of the effect, we chose to keep the height, eccentricity, and full width half maximum diameter consistent and vary only the taper angle. A schematic of the micropillar and the varying taper angle is presented in Figure S6. Parameterising the model with the median micromechanical properties and varying the taper angle between the measured maximum and minimum taper angles of 10.4° and 14°, accounted for approximately 50.6% and 106.9% of the IQR of $K_{app}$ and $F_Y$ respectively. We speculate that the more pronounced effect on $F_Y$ is driven by the reduced area of contact with the indenter tip, resulting in an increased stress concentration at the contact interface [68]. The asymmetry between the impact on numerical $K_{app}$ and $F_Y$, and the consistent variation between the experimental $K_{app}$ and $F_Y$ may suggest that the latter is driven by other factors like lamella anisotropy. Varying the micropillar geometry, however, has highlighted the importance of a highly repeatable micropillar extraction technique to minimise the impact of geometric variability. Tailoring the *in silico* modelling to include the pillar specific geometry could minimise this impact, however, it would require additional manual input and compromise the benefit of the automated minimisation routine developed to characterise the micromechanical properties of the constituent material.

The variation of material orientation explains the large variation in observed here. The large number of nanoindentation measurements was sufficient to detect a small (0.81 GPa, 95% CI [0.05, 1.58]) but significant increase in means between ND and OA subchondral trabecular bone. Given that stiffnesses measured by micropillar compression and nanoindentation in this study fit well together (Figure 5), this difference may persist when testing via micropillar compression. However, 109 micropillars may have been too few to detect a significant difference in $E_0$ given the large variation. We did detect a significant difference in yield strength between ND and OA subchondral trabecular bone (Table 1). As with the nanoindentation measurement locations, micropillars were positioned centrally within junctures of trabeculae (Figure 8b). We believe that the increased tissue maturity in the centre of trabeculae is relevant here as well as in the nanoindentation measurements, given the surface deposition of newly formed bone. Ultimately, such small changes in tissue level stiffness and strength may be overshadowed by the densification of the subchondral bone in late-stage OA [51]. With marginal changes to tissue-scale properties, we may infer that any changes to apparent level stiffness and strength are a consequence of microstructural changes. For example, Stadelmann et al. [69] illustrate this nicely, where a reduction in tissue scale stiffness in osteoblastic vertebral lesions is inconsequential with respect to apparent level stiffness due to an increase in overall bone mass.

## 4.3. Compositional analysis

The pillar-specific Raman spectroscopic metrics we evaluated exhibited ranges that are comparable to previous studies on bone tissue [22,28,36,37]. Despite this, the only location-based significant difference was that the mineral crystallinity was reduced in the OA subchondral bone plate. This finding is in contrast to Das Gupta et al. [70] who performed Raman mapping on the osteochondral junction from proximal tibiae with progressive stages of OA and found no change in the mineral crystallinity in the subchondral bone plate. Mineral crystallinity, which encompasses both crystal size and perfection, increases with tissue age [71,72]. In the context of tissue age, our detected reduced mineral crystallinity in the OA subchondral bone plate fits the notion of newly formed bone responsible for bone plate thickening observed in OA progression [51]. Mineral crystallinity has previously been shown to correlate with elasticity and yield strength [37,71,73], and when the OA subchondral bone plate and trabeculae elastic modulus and yield strength results were pooled, we detected a significant dependence of both on mineral crystallinity (Figure 6). In the ND bone, the mineralisation density derived by qBEI was the dominant explanatory variable (Figure 6). This mineralisation density dependence is consistent with previous experimental and numerical observations of micropillar compression [21,40]. Despite the detection of reduced crystallinity between location, the chosen Raman metrics related to the collagen conformation in bone proved no different between the OA and ND groups nor location. Interestingly $I_{1670}/I_{1640}$ was identified as a significant explanatory variable of $E_0$ in the power law regression analysis in the bone plate of ND specimens. This metric has previously only been associated with a loss of mechanical competence, through reduced fracture toughness and strength [37,73].

The lack of groupwise changes could have been a sensitivity issue with the chosen fitting method of the Amide I band. Two different fitting procedures could be used and both achieve a high coefficient of determination, but the derived ratios may not correlate (see [73]). It could also simply be that the collagen conformation in the tissue that we probe is unaffected by OA. The positioning bias, where we could not probe the hypomineralised lamella packets, could extend to the Raman analysis as the measurements were spatially matched to the micropillars. As with the mechanical testing, we may have detected changes in collagen conformation had we been able to probe these regions.

## 4.4. Limitations

The male-only nature of the study was an outcome of the types of patients undergoing arthroplasty at the time of tissue collection and not of study design. ND specimens were then sex-matched to OA ones.

The necessity to back-calculate the micromechanical properties through an *in silico* experiment highlights the limitations of our current laser ablation protocol. Our achieved pillar geometry prohibits an analytical transformation of the force-displacement data directly to a uniaxial stress-strain curve. The limiting factor here is the conical shape of the pillars, which induces a heterogeneous stress state within the pillar under compression [68]. We believe that the modelling approach sufficiently captures the elastic and post-yield behaviour of the material that comprises each pillar. Ideally, further effort needs to be conducted to deliver near cylindrical pillars by laser ablation, for example like those produced by Lim et al [74]. However, if these are achieved height needs to be controlled to reduce the risk of buckling. In our case the top diameter to height ratio would have been too small to securely avoid this and shorter pillars would then also be needed.

The positioning of the pillars in trabeculae was a challenge as we had no volumetric spatial information on the bone tissue beneath the surface of the sample. With only a light microscope surface map, our plan was to fabricate a large number of pillars so that a sufficient number would be suitable for testing. Numerous pillars were rejected due to insufficient bone volume beneath the surface or pores. On reflection, this was to be expected given the size of the pillars. While we selected positions that suggested sufficient material, the number of rejected pillars indicates that they were on the upper limit of what was possible. Along with eliminating the taper angle, the size of pillars should therefore be reduced to improve the fabrication success rate.

Due to the pillar size and indentation depth, no measurements could be located in hypomineralised lamella. These regions were inaccessible by means of micropillar compression or our nanoindentation protocol as they were located on peripheries of trabeculae (see Figure 2b). Micropillars with a diameter of approximately 5 µm can be achieved by focused ion beam milling [19,21,25,26,75], which could be small enough to probe these areas. However, as trabecular have a round cross section, peripheral regions would provide too little support material underneath the probed bone. Li and Aspden [76] illustrated that OA trabecular bone exhibits a significantly lower apparent stiffness than non-diseased bone with comparable apparent density. Increases in apparent density in OA is accommodated by an increase in bone volume with reduced tissue mineralisation [77]. Considering our detected mineralisation dependence of the micromechanical properties, we speculate that these hypomineralised regions may explain the density associated loss of apparent stiffness demonstrated by Li and Aspden [76]. Interestingly, Das Gupta et al. [78] performed microindentations in arrays across the osteochondral junction and found a loss of tissue stiffness with progressive cartilage degeneration. This systematic approach avoids the need to subjectively locate measurements. In our case, we chose the location of the pillars due to their size and ultimately could not test the full range of mineralisation heterogeneity. Systematic positioning, as demonstrated by Kochetkova et al. [22,28], could overcome this problem as micropillars would be randomly distributed across the bone sample. However, Kochetkova et al. [22,28] did not experience the challenge of trabecular porosity as their pillars were extracted in cortical bone, where intracortical porosity is a fraction of that in trabecular bone.

# 5. Conclusion

We used nanoindentation and micropillar compression to investigate the micromechanical properties of late-stage OA subchondral bone. This combination allowed us to observe tissue yield strength in addition to tissue stiffness, which to the best of our knowledge remained an open question and constitutes a key outcome of this study. Nanoindentation revealed a small but significant increase in elastic modulus, and while the elastic modulus derived by micropillar compression exhibited a comparable range, no significant differences were detected between ND and OA bone. Both testing modes showed large variation, and we consider the impact of lamellar anisotropy to be the likely source of this variability. We detected a significant difference in yield strength between ND and OA subchondral trabecular bone that was absent in the subchondral bone plate. Mineralization proved to be a significant determinant of elastic modulus and compressive yield strength, confirming its dominant role in these properties.

The results provide insight into subtle tissue-level changes that occur alongside drastic apparent level changes in OA. The morphology of samples included in our study was previously evaluated by Koria et al. [15], who reported a >10% increase in bone volume fraction with OA. Given that bone volume fraction is the primary determinant of apparent level stiffness and strength, an estimated increase in elastic modulus of only 0.81 GPa in the subchondral trabeculae may be inconsequential. However, these subtle tissue-level changes may suggest that subchondral trabecular bone could assist with localised stress absorption leading to pathological bone remodelling [14,79] that would trigger bone cells to produce greater amount of bone [76]. Our findings contribute new insights into the micromechanical properties of OA subchondral bone, which may be used directly in the development of computational methods for assessing subchondral bone stiffness and strength, for example for use in surgical planning or implant design or to inform OA therapeutic approached by targeting subchondral bone mechanobiology.

## Declaration of Competing Interest

The authors declare that they have no known competing financial interests or personal relationships that could have appeared to influence the work reported in this paper.

## Acknowledgements

This work was supported by the Engineering and Physical Sciences Research Council (EPSRC), UK (Grant EP/P005756/1) and Leverhulme Trust Research Project Grant (RPG-2020-215).

## References

[1] U.T. Kadam, Clinical comorbidity in patients with osteoarthritis: a case-control study of general practice consulters in England and Wales, Annals of the Rheumatic Diseases 63 (2004) 408–414. https://doi.org/10.1136/ard.2003.007526.

[2] T. Neogi, The epidemiology and impact of pain in osteoarthritis, Osteoarthritis and Cartilage 21 (2013) 1145–1153. https://doi.org/10.1016/j.joca.2013.03.018.

[3] D.J. Hunter, D. Schofield, E. Callander, The individual and socioeconomic impact of osteoarthritis, Nat Rev Rheumatol 10 (2014) 437–441. https://doi.org/10.1038/nrrheum.2014.44.

[4] J.D. Steinmetz, G.T. Culbreth, L.M. Haile, Q. Rafferty, J. Lo, K.G. Fukutaki, J.A. Cruz, A.E. Smith, S.E. Vollset, P.M. Brooks, M. Cross, A.D. Woolf, H. Hagins, M. Abbasi-Kangevari, A. Abedi, I.N. Ackerman, H. Amu, B. Antony, J. Arabloo, A.Y. Aravkin, A.M. Argaw, A.A. Artamonov, T. Ashraf, A. Barrow, L.M. Bearne, I.M. Bensenor, A.Y. Berhie, N. Bhardwaj, P. Bhardwaj, V.S. Bhojaraja, A. Bijani, P.S. Briant, A.M. Briggs, N.S. Butt, J. Charan, V.K. Chattu, F.M. Cicuttini, K. Coberly, O. Dadras, X. Dai, L. Dandona, R. Dandona, K. de Luca, E. Denova-Gutiérrez, S.D. Dharmaratne, M. Dhimal, M. Dianatinasab, K.E. Dreinhoefer, M. Elhadi, U. Farooque, H.R. Farpour, I. Filip, F. Fischer, M. Freitas, B. Ganesan, B.N.B. Gemeda, T. Getachew, S.H. Ghamari, A. Ghashghaee, T.K. Gill, M. Golechha, D. Golinelli, B. Gupta, V.B. Gupta, V.K. Gupta, R. Haddadi, N. Hafezi-Nejad, R. Halwani, S. Hamidi, A. Hanif, N.I. Harlianto, J.M. Haro, J. Hartvigsen, S.I. Hay, J.J. Hebert, G. Heidari, M.S. Hosseini, M. Hosseinzadeh, A.K. Hsiao, I.M.

Ilic, M.D. Ilic, L. Jacob, R. Jayawardena, R.P. Jha, J.B. Jonas, N. Joseph, H. Kandel, I.M. Karaye, M.J. Khan, Y.J. Kim, A.A. Kolahi, O. Korzh, R. Koteeswaran, V. Krishnamoorthy, G.A. Kumar, N. Kumar, S.W. Lee, S.S. Lim, S.W. Lobo, G. Lucchetti, M.R. Malekpour, A.A. Malik, L.G.G. Mandarano-Filho, S. Martini, A.F.A. Mentis, M.K. Mesregah, T. Mestrovic, E.M. Mirrakhimov, A. Misganaw, R. Mohammadpourhodki, A.H. Mokdad, S. Momtazmanesh, S.D. Morrison, C.J.L. Murray, H. Nassereldine, H.B. Netsere, S.N. Kandel, M.O. Owolabi, S. Panda-Jonas, A. Pandey, S. Pawar, P. Pedersini, J. Pereira, A. Radfar, M.M. Rashidi, D.L. Rawaf, S. Rawaf, R. Rawassizadeh, S.M. Rayegani, D. Ribeiro, L. Roever, B. Saddik, A. Sahebkar, S. Salehi, L.S. Riera, F. Sanmarchi, M.M. Santric-Milicevic, S. Shahabi, M.A. Shaikh, E. Shaker, M. Shannawaz, R. Sharma, S. Sharma, J.K. Shetty, R. Shiri, P. Shobeiri, D.A.S. Silva, A. Singh, J.A. Singh, S. Singh, S.T. Skou, H. Slater, M.S. Soltani-Zangbar, A.V. Starodubova, A. Tehrani-Banihashemi, S.V. Tahbaz, P.R. Valdez, B. Vo, L.G. Vu, Y.P. Wang, S.H.Y. Jabbari, N. Yonemoto, I. Yunusa, L.M. March, K.L. Ong, T. Vos, J.A. Kopec, Global, regional, and national burden of osteoarthritis, 1990-2020 and projections to 2050: a systematic analysis for the Global Burden of Disease Study 2021, The Lancet Rheumatology 5 (2023) e508–e522. https://doi.org/10.1016/S2665-9913(23)00163-7.

[5] Y. Cho, S. Jeong, H. Kim, D. Kang, J. Lee, S.-B. Kang, J.-H. Kim, Disease-modifying therapeutic strategies in osteoarthritis: current status and future directions, Exp Mol Med 53 (2021) 1689–1696. https://doi.org/10.1038/s12276-021-00710-y.

[6] N.K. Arden, T.A. Perry, R.R. Bannuru, O. Bruyère, C. Cooper, I.K. Haugen, M.C. Hochberg, T.E. McAlindon, A. Mobasheri, J.-Y. Reginster, Non-surgical management of knee osteoarthritis: comparison of ESCEO and OARSI 2019 guidelines, Nat Rev Rheumatol 17 (2021) 59–66. https://doi.org/10.1038/s41584-020-00523-9.

[7] L.A. Deveza, A.E. Nelson, R.F. Loeser, Phenotypes of osteoarthritis: current state and future implications, Clinical and Experimental Rheumatology 37 (2019) 64–72.

[8] M. Herrera-Pérez, V. Valderrabano, A.L. Godoy-Santos, C. de César Netto, D. González-Martín, S. Tejero, Ankle osteoarthritis: comprehensive review and treatment algorithm proposal, EFORT Open Reviews 7 (2022) 448–459. https://doi.org/10.1530/EOR-21-0117.

[9] A.S. Bawa, D.T. Loveday, Intra-articular osteotomy for ankle arthritis, Orthopaedics and Trauma 37 (2023) 34–39. https://doi.org/10.1016/j.mporth.2022.11.005.

[10] Y. Yasui, C.P. Hannon, D. Seow, J.G. Kennedy, Ankle arthrodesis: A systematic approach and review of the literature, WJO 7 (2016) 700. https://doi.org/10.5312/wjo.v7.i11.700.

[11] L.W. van der Plaat, D. Hoornenborg, I.N. Sierevelt, C.N. van Dijk, D. Haverkamp, Ten-year revision rates of contemporary total ankle arthroplasties equal 22%. A meta-analysis, Foot and Ankle Surgery 28 (2022) 543–549. https://doi.org/10.1016/j.fas.2021.05.014.

[12] M.S. Lee, L. Mathson, C. Andrews, D. Wiese, J.M. Fritz, A.E. Jimenez, B. Law, Long-term Outcomes After Total Ankle Arthroplasty: A Systematic Review, Foot and Ankle Orthopaedics 9 (2024). https://doi.org/10.1177/24730114241294073.

[13] C. Egloff, T. Hügle, V. Valderrabano, Biomechanics and pathomechanisms of osteoarthritis, Swiss Medical Weekly 142 (2012) 1–14. https://doi.org/10.4414/smw.2012.13583.

[14] R. Feng, W. Hu, Y. Li, X. Yao, J. Li, X. Li, J. Zhang, Y. Wu, F. Kang, S. Dong, Mechanotransduction in subchondral bone microenvironment and targeted interventions for osteoarthritis, Mechanobiology in Medicine 2 (2024). https://doi.org/10.1016/j.mbm.2024.100043.

[15] L. Koria, M. Farndon, E. Jones, M. Mengoni, C. Brockett, Changes in subchondral bone morphology with osteoarthritis in the ankle, PLoS ONE 19 (2024) e0290914. https://doi.org/10.1371/journal.pone.0290914.

[16] J.B. Renault, M. Carmona, C. Tzioupis, M. Ollivier, J.N. Argenson, S. Parratte, P. Chabrand, Tibial subchondral trabecular bone micromechanical and microarchitectural properties are affected by alignment and osteoarthritis stage, Scientific Reports 10 (2020) 1–10. https://doi.org/10.1038/s41598-020-60464-x.

[17] A.E. Peters, R. Akhtar, E.J. Comerford, K.T. Bates, The effect of ageing and osteoarthritis on the mechanical properties of cartilage and bone in the human knee joint, Scientific Reports 8 (2018) 1–13. https://doi.org/10.1038/s41598-018-24258-6.

[18] Q. Zuo, S. Lu, Z. Du, T. Friis, J. Yao, R. Crawford, I. Prasadam, Y. Xiao, Characterization of nano-structural and nano-mechanical properties of osteoarthritic subchondral bone, BMC Musculoskeletal Disorders 17 (2016) 1–13. https://doi.org/10.1186/s12891-016-1226-1.

[19] J. Schwiedrzik, A. Taylor, D. Casari, U. Wolfram, P. Zysset, J. Michler, Nanoscale deformation mechanisms and yield properties of hydrated bone extracellular matrix, Acta Biomaterialia 60 (2017) 302–314. https://doi.org/10.1016/j.actbio.2017.07.030.

[20] A. Groetsch, A. Gourrier, D. Casari, J. Schwiedrzik, J.D. Shephard, J. Michler, P.K. Zysset, U. Wolfram, The elasto-plastic nano- and microscale compressive behaviour of rehydrated mineralised collagen fibres, Acta Biomaterialia 164 (2023) 332–345. https://doi.org/10.1016/j.actbio.2023.03.041.

[21] M. Indermaur, D. Casari, T. Kochetkova, C. Peruzzi, E. Zimmermann, F. Rauch, B. Willie, J. Michler, J. Schwiedrzik, P. Zysset, Compressive Strength of Iliac Bone ECM Is Not Reduced in Osteogenesis Imperfecta and Increases With Mineralization, Journal of Bone and Mineral Research 36 (2021) 1364–1375. https://doi.org/10.1002/jbmr.4286.

[22] T. Kochetkova, M.S. Hanke, M. Indermaur, A. Groetsch, S. Remund, B. Neuenschwander, J. Michler, K.A. Siebenrock, P. Zysset, J. Schwiedrzik, Composition and micromechanical properties of the femoral neck compact bone in relation to patient age, sex and hip fracture occurrence, Bone 177 (2023) 116920. https://doi.org/10.1016/j.bone.2023.116920.

[23] C. Peruzzi, R. Ramachandramoorthy, A. Groetsch, D. Casari, P. Grönquist, M. Rüggeberg, J. Michler, J. Schwiedrzik, Microscale compressive behavior of hydrated lamellar bone at high strain rates, Acta Biomaterialia 131 (2021) 403–414. https://doi.org/10.1016/j.actbio.2021.07.005.

[24] T. Kochetkova, C. Peruzzi, O. Braun, J. Overbeck, A.K. Maurya, A. Neels, M. Calame, J. Michler, P. Zysset, J. Schwiedrzik, Combining polarized Raman spectroscopy and micropillar compression to study microscale structure-property relationships in mineralized tissues, Acta Biomaterialia 119 (2021) 390–404. https://doi.org/10.1016/j.actbio.2020.10.034.

[25] J. Schwiedrzik, R. Raghavan, A. Bürki, V. Lenader, U. Wolfram, J. Michler, P. Zysset, In situ micropillar compression reveals superior strength and ductility but an absence of damage in lamellar bone, Nature Materials 13 (2014) 740–747. https://doi.org/10.1038/nmat3959.

[26] A. Groetsch, A. Gourrier, J. Schwiedrzik, M. Sztucki, R.J. Beck, J.D. Shephard, J. Michler, P.K. Zysset, U. Wolfram, Compressive behaviour of uniaxially aligned individual mineralised collagen fibres at the micro- and nanoscale, Acta Biomaterialia 89 (2019) 313–329. https://doi.org/10.1016/j.actbio.2019.02.053.

[27] U. Wolfram, M. Peña Fernández, S. McPhee, E. Smith, R.J. Beck, J.D. Shephard, A. Ozel, C.S. Erskine, J. Büscher, J. Titschack, J.M. Roberts, S.J. Hennige, Multiscale mechanical consequences of ocean acidification for cold-water corals, Sci Rep 12 (2022) 8052. https://doi.org/10.1038/s41598-022-11266-w.

[28] T. Kochetkova, A. Groetsch, M. Indermaur, C. Peruzzi, S. Remund, B. Neuenschwander, B. Bellon, J. Michler, P. Zysset, J. Schwiedrzik, Assessing minipig compact jawbone quality at the microscale, Journal of the Mechanical Behavior of Biomedical Materials 134 (2022) 105405. https://doi.org/10.1016/j.jmbbm.2022.105405.

[29] M.D. Kohn, A.A. Sassoon, N.D. Fernando, Classifications in Brief: Kellgren-Lawrence Classification of Osteoarthritis, Clinical Orthopaedics and Related Research 474 (2016) 1886–1893. https://doi.org/10.1007/s11999-016-4732-4.

[30] M.J. Mirzaali, J.J. Schwiedrzik, S. Thaiwichai, J.P. Best, J. Michler, P.K. Zysset, U. Wolfram, Mechanical properties of cortical bone and their relationships with age, gender, composition and microindentation properties in the elderly, Bone 93 (2016) 196–211. https://doi.org/10.1016/j.bone.2015.11.018.

[31] U. Wolfram, H.-J. Wilke, P. Zysset, Transverse Isotropic elastic properties of vertebral trabecular bone matrix measured using microindentation under dry conditions (Effects of Age, Gender and Vertebral Level), Journal of Mechanics in Medicine and Biology 10 (2010) 139–150. https://doi.org/10.1142/S0219519410003241.

[32] P. Roschger, P. Fratzl, J. Eschberger, K. Klaushofer, Validation of quantitative backscattered electron imaging for the measurement of mineral density distribution in human bone biopsies, Bone 23 (1998) 319–326. https://doi.org/10.1016/S8756-3282(98)00112-4.

[33] M.D. Abràmoff, P.J. Magalhães, S.J. Ram, Image processing with imageJ, Biophotonics International 11 (2004) 36–41. https://doi.org/10.1201/9781420005615.ax4.

[34] D. Erb, pybaselines: A Python library of algorithms for the baseline correction of experimental data, (2022). https://doi.org/10.5281/ZENODO.7255880.

[35] P.H.C. Eilers, A Perfect Smoother, Anal. Chem. 75 (2003) 3631–3636. https://doi.org/10.1021/ac034173t.

[36] M. Unal, H. Jung, O. Akkus, Novel Raman Spectroscopic Biomarkers Indicate That Postyield Damage Denatures Bone’s Collagen, Journal of Bone and Mineral Research 31 (2016) 1015–1025. https://doi.org/10.1002/jbmr.2768.

[37] M. Unal, S. Uppuganti, S. Timur, A. Mahadevan-Jansen, O. Akkus, J.S. Nyman, Assessing matrix quality by Raman spectroscopy helps predict fracture toughness of human cortical bone, Scientific Reports 9 (2019) 1–13. https://doi.org/10.1038/s41598-019-43542-7.

[38] W.C. Oliver, G.M. Pharr, An improved technique for determining hardness and elastic modulus using load and displacement sensing indentation experiments, J. Mater. Res. 7 (1992) 1564–1583. https://doi.org/10.1557/JMR.1992.1564.

[39] A.C. Fischer-Cripps, Critical review of analysis and interpretation of nanoindentation test data, Surface and Coatings Technology 200 (2006) 4153–4165. https://doi.org/10.1016/j.surfcoat.2005.03.018.

[40] A. Groetsch, P.K. Zysset, P. Varga, A. Pacureanu, F. Peyrin, U. Wolfram, An experimentally informed statistical elasto-plastic mineralised collagen fibre model at the micrometre and nanometre lengthscale, Sci Rep 11 (2021) 15539. https://doi.org/10.1038/s41598-021-93505-0.

[41] M. Indermaur, D. Casari, T. Kochetkova, C. Peruzzi, E. Zimmermann, F. Rauch, B. Willie, J. Michler, J. Schwiedrzik, P. Zysset, Compressive Strength of Iliac Bone ECM Is Not Reduced in Osteogenesis Imperfecta and Increases With Mineralization, Journal of Bone and Mineral Research 36 (2021) 1364–1375. https://doi.org/10.1002/jbmr.4286.

[42] H. Zhang, B.E. Schuster, Q. Wei, K.T. Ramesh, The design of accurate micro-compression experiments, Scripta Materialia 54 (2006) 181–186. https://doi.org/10.1016/j.scriptamat.2005.06.043.

[43] J. Schwiedrzik, T. Gross, M. Bina, M. Pretterklieber, P. Zysset, D. Pahr, Experimental validation of a nonlinear $\mu$ FE model based on cohesive-frictional plasticity for trabecular bone, Numer Methods Biomed Eng 32 (2016) e02739. https://doi.org/10.1002/cnm.2739.

[44] J.J. Schwiedrzik, U. Wolfram, P.K. Zysset, A generalized anisotropic quadric yield criterion and its application to bone tissue at multiple length scales, Biomechanics and Modeling in Mechanobiology 12 (2013) 1155–1168. https://doi.org/10.1007/s10237-013-0472-5.

[45] M.J. Mirzaali, A. Bürki, J. Schwiedrzik, P.K. Zysset, U. Wolfram, Continuum damage interactions between tension and compression in osteonal bone, Journal of the Mechanical Behavior of Biomedical Materials 49 (2015) 355–369. https://doi.org/10.1016/j.jmbbm.2015.05.007.

[46] U. Wolfram, T. Gross, D.H. Pahr, J. Schwiedrzik, H.J. Wilke, P.K. Zysset, Fabric-based Tsai-Wu yield criteria for vertebral trabecular bone in stress and strain space, Journal of the Mechanical Behavior of Biomedical Materials 15 (2012) 218–228. https://doi.org/10.1016/j.jmbbm.2012.07.005.

[47] D. Casari, J. Michler, P. Zysset, J. Schwiedrzik, Microtensile properties and failure mechanisms of cortical bone at the lamellar level, Acta Biomaterialia 120 (2021) 135–145. https://doi.org/10.1016/j.actbio.2020.04.030.

[48] S. Seabold, J. Perktold, Statsmodels: Econometric and Statistical Modeling with Python, Proceedings of the 9th Python in Science Conference (2010) 92–96. https://doi.org/10.25080/majora-92bf1922-011.

[49] R. Lenth, emmeans: Estimated marginal means, aka least-squares means., R package version 1.7. 2. (2022).

[50] U. Wolfram, H.J. Wilke, P.K. Zysset, Rehydration of vertebral trabecular bone: Influences on its anisotropy, its stiffness and the indentation work with a view to age, gender and vertebral level, Bone 46 (2010) 348–354. https://doi.org/10.1016/j.bone.2009.09.035.

[51] M.A.J. Finnilä, J. Thevenot, O.M. Aho, V. Tiitu, J. Rautiainen, S. Kauppinen, M.T. Nieminen, K. Pritzker, M. Valkealahti, P. Lehenkari, S. Saarakkala, Association between subchondral bone structure and osteoarthritis histopathological grade, Journal of Orthopaedic Research 35 (2017) 785–792. https://doi.org/10.1002/jor.23312.

[52] S. Hengsberger, A. Kulik, P. Zysset, Nanoindentation discriminates the elastic properties of individual human bone lamellae under dry and physiological conditions, Bone 30 (2002) 178–184. https://doi.org/10.1016/S8756-3282(01)00624-X.

[53] G. Guidoni, M. Swain, I. Jäger, Nanoindentation of wet and dry compact bone: Influence of environment and indenter tip geometry on the indentation modulus, Philosophical Magazine 90 (2010) 553–565. https://doi.org/10.1080/14786430903201853.

[54] P.K. Zysset, X. Edward Guo, C. Edward Hoffler, K.E. Moore, S.A. Goldstein, Elastic modulus and hardness of cortical and trabecular bone lamellae measured by nanoindentation in the human femur, Journal of Biomechanics 32 (1999) 1005–1012. https://doi.org/10.1016/S0021-9290(99)00111-6.

[55] D. Casari, T. Kochetkova, J. Michler, P. Zysset, J. Schwiedrzik, Microtensile failure mechanisms in lamellar bone: Influence of fibrillar orientation, specimen size and hydration, Acta Biomaterialia 131 (2021) 391–402. https://doi.org/10.1016/j.actbio.2021.06.032.

[56] H.H. Bayraktar, E.F. Morgan, G.L. Niebur, G.E. Morris, E.K. Wong, T.M. Keaveny, Comparison of the elastic and yield properties of human femoral trabecular and cortical bone tissue, Journal of Biomechanics 37 (2004) 27–35. https://doi.org/10.1016/S0021-9290(03)00257-4.

[57] O.A. Tertuliano, J.R. Greer, The nanocomposite nature of bone drives its strength and damage resistance, Nature Mater 15 (2016) 1195–1202. https://doi.org/10.1038/nmat4719.

[58] M. Frank, A.G. Reisinger, D.H. Pahr, P.J. Thurner, Effects of Osteoporosis on Bone Morphometry and Material Properties of Individual Human Trabeculae in the Femoral Head, JBMR Plus 5 (2021) e10503. https://doi.org/10.1002/jbm4.10503.

[59] P. Zysset, A constitutive law for trabecular bone, 1252 (1994) 231.

[60] U. Wolfram, J. Schwiedrzik, Post-yield and failure properties of cortical bone, BoneKEy Reports 5 (2016) 1–10. https://doi.org/10.1038/bonekey.2016.60.

[61] H.H. Bayraktar, T.M. Keaveny, Mechanisms of uniformity of yield strains for trabecular bone, Journal of Biomechanics 37 (2004) 1671–1678. https://doi.org/10.1016/j.jbiomech.2004.02.045.

[62] P. Roschger, E.P. Paschalis, P. Fratzl, K. Klaushofer, Bone mineralization density distribution in health and disease, Bone 42 (2008) 456–466. https://doi.org/10.1016/j.bone.2007.10.021.

[63] E. Goff, F. Buccino, C. Bregoli, J.P. McKinley, B. Aeppli, R.R. Recker, E. Shane, A. Cohen, G. Kuhn, R. Müller, Large-scale quantification of human osteocyte lacunar morphological biomarkers as assessed by ultra-high-resolution desktop micro-computed tomography, Bone 152 (2021) 116094. https://doi.org/10.1016/j.bone.2021.116094.

[64] P. Virtanen, R. Gommers, T.E. Oliphant, M. Haberland, T. Reddy, D. Cournapeau, E. Burovski, P. Peterson, W. Weckesser, J. Bright, S.J. van der Walt, M. Brett, J. Wilson, K.J. Millman, N. Mayorov, A.R.J. Nelson, E. Jones, R. Kern, E. Larson, C.J. Carey, İ. Polat, Y. Feng, E.W. Moore, J. VanderPlas, D. Laxalde, J. Perktold, R. Cimrman, I. Henriksen, E.A. Quintero, C.R. Harris, A.M. Archibald, A.H. Ribeiro, F. Pedregosa, P. van Mulbregt, SciPy 1.0 Contributors, A. Vijaykumar, A.P. Bardelli, A. Rothberg, A. Hilboll, A. Kloeckner, A. Scopatz, A. Lee, A. Rokem, C.N. Woods, C. Fulton, C. Masson, C. Häggström, C. Fitzgerald, D.A. Nicholson, D.R. Hagen, D.V. Pasechnik, E. Olivetti, E. Martin, E. Wieser, F. Silva, F. Lenders, F. Wilhelm, G. Young, G.A. Price, G.-L. Ingold, G.E. Allen, G.R. Lee, H. Audren, I. Probst, J.P. Dietrich, J. Silterra, J.T. Webber, J. Slavič, J. Nothman, J. Buchner, J. Kulick, J.L. Schönberger, J.V. de Miranda Cardoso, J. Reimer, J. Harrington, J.L.C. Rodríguez, J. Nunez-Iglesias, J. Kuczynski, K. Tritz, M.

Thoma, M. Newville, M. Kümmerer, M. Bolingbroke, M. Tartre, M. Pak, N.J. Smith, N. Nowaczyk, N. Shebanov, O. Pavlyk, P.A. Brodtkorb, P. Lee, R.T. McGibbon, R. Feldbauer, S. Lewis, S. Tygier, S. Sievert, S. Vigna, S. Peterson, S. More, T. Pudlik, T. Oshima, T.J. Pingel, T.P. Robitaille, T. Spura, T.R. Jones, T. Cera, T. Leslie, T. Zito, T. Krauss, U. Upadhyay, Y.O. Halchenko, Y. Vázquez-Baeza, SciPy 1.0: fundamental algorithms for scientific computing in Python, Nat Methods 17 (2020) 261–272. https://doi.org/10.1038/s41592-019-0686-2.

[65] A.G. Reisinger, D.H. Pahr, P.K. Zysset, Elastic anisotropy of bone lamellae as a function of fibril orientation pattern, Biomechanics and Modeling in Mechanobiology 10 (2011) 67–77. https://doi.org/10.1007/s10237-010-0218-6.

[66] P. Varga, A. Pacureanu, M. Langer, H. Suhonen, B. Hesse, Q. Grimal, P. Cloetens, K. Raum, F. Peyrin, Investigation of the three-dimensional orientation of mineralized collagen fibrils in human lamellar bone using synchrotron X-ray phase nano-tomography, Acta Biomaterialia 9 (2013) 8118–8127. https://doi.org/10.1016/j.actbio.2013.05.015.

[67] A. Speed, A. Groetsch, J.J. Schwiedrzik, U. Wolfram, Extrafibrillar matrix yield stress and failure envelopes for mineralised collagen fibril arrays, Journal of the Mechanical Behavior of Biomedical Materials 105 (2020) 103563. https://doi.org/10.1016/j.jmbbm.2019.103563.

[68] R. Lacroix, G. Kermouche, J. Teisseire, E. Barthel, Plastic deformation and residual stresses in amorphous silica pillars under uniaxial loading, Acta Materialia 60 (2012) 5555–5566. https://doi.org/10.1016/j.actamat.2012.07.016.

[69] M.A. Stadelmann, D.E. Schenk, G. Maquer, C. Lenherr, F.M. Buck, D.D. Bosshardt, S. Hoppe, N. Theumann, R.N. Alkalay, P.K. Zysset, Conventional finite element models estimate the strength of metastatic human vertebrae despite alterations of the bone's tissue and structure, Bone 141 (2020) 115598. https://doi.org/10.1016/j.bone.2020.115598.

[70] S. Das Gupta, M.A.J. Finnilä, S.S. Karhula, S. Kauppinen, A. Joukainen, H. Kröger, R.K. Korhonen, A. Thambyah, L. Rieppo, S. Saarakkala, Raman microspectroscopic analysis of the tissue-specific composition of the human osteochondral junction in osteoarthritis: A pilot study, Acta Biomaterialia 106 (2020) 145–155. https://doi.org/10.1016/j.actbio.2020.02.020.

[71] J.S. Yerramshetty, O. Akkus, The associations between mineral crystallinity and the mechanical properties of human cortical bone, Bone 42 (2008) 476–482. https://doi.org/10.1016/j.bone.2007.12.001.

[72] J.J. Freeman, B. Wopenka, M.J. Silva, J.D. Pasteris, Raman Spectroscopic Detection of Changes in Bioapatite in Mouse Femora as a Function of Age and In Vitro Fluoride Treatment, Calcif Tissue Int 68 (2001) 156–162. https://doi.org/10.1007/s002230001206.

[73] M. Unal, Raman spectroscopic determination of bone matrix quantity and quality augments prediction of human cortical bone mechanical properties, Journal of Biomechanics 119 (2021) 110342. https://doi.org/10.1016/j.jbiomech.2021.110342.

[74] Y.C. Lim, K. Altman, D.F. Farson, K.M. Flores, Micropillar fabrication on bovine cortical bone by direct write femtosecond laser ablation, 29th International Congress on Applications of Lasers and Electro-Optics, ICALEO 2010 - Congress Proceedings 103 (2010) 1124–1132. https://doi.org/10.1117/1.3268444.

[75] K.W. Luczynski, A. Steiger-Thirsfeld, J. Bernardi, J. Eberhardsteiner, C. Hellmich, Extracellular bone matrix exhibits hardening elastoplasticity and more than double cortical strength: Evidence from homogeneous compression of non-tapered single micron-sized pillars welded to a rigid substrate, Journal of the Mechanical Behavior of Biomedical Materials 52 (2015) 51–62. https://doi.org/10.1016/j.jmbbm.2015.03.001.

[76] B. Li, R.M. Aspden, Composition and Mechanical Properties of Cancellous Bone from the Femoral Head of Patients with Osteoporosis or Osteoarthritis, J Bone Miner Res 12 (1997) 641–651. https://doi.org/10.1359/jbmr.1997.12.4.641.

[77] L.G.E. Cox, C.C. van Donkelaar, B. van Rietbergen, P.J. Emans, K. Ito, Decreased bone tissue mineralization can partly explain subchondral sclerosis observed in osteoarthritis, Bone 50 (2012) 1152–1161. https://doi.org/10.1016/j.bone.2012.01.024.

[78] S. Das Gupta, J. Workman, M.A.J. Finnilä, S. Saarakkala, A. Thambyah, Subchondral bone plate thickness is associated with micromechanical and microstructural changes in the bovine patella osteochondral junction with different levels of cartilage degeneration, Journal of the Mechanical Behavior of Biomedical Materials 129 (2022) 105158. https://doi.org/10.1016/j.jmbbm.2022.105158.

[79] Q. Zuo, S. Lu, Z. Du, T. Friis, J. Yao, R. Crawford, I. Prasadam, Y. Xiao, Characterization of nano-structural and nano-mechanical properties of osteoarthritic subchondral bone, BMC Musculoskelet Disord 17 (2016) 367. https://doi.org/10.1186/s12891-016-1226-1.

Supplementary material to:

# Micromechanical characterisation of osteoarthritic subchondral bone by micropillar compression

Samuel McPhee[1], Marta Peña Fernández[1], Lekha Koria[2], Marlène Mengoni[2], Rainer J Beck[3], Jonathan D Shephard[3], Claire Brockett[4], Uwe Wolfram[1,5*]

[1] Institute of Mechanical, Process and Energy Engineering, Heriot-Watt University, UK.
[2] Institute of Medical and Biological Engineering, University of Leeds, UK.
[3] Institute of Photonics and Quantum Sciences, Heriot-Watt University, UK.
[4] INSIGNEO Institute for in silico Medicine, University of Sheffield, UK.
[5] Institute for Material Science and Engineering, TU Clausthal, Germany.

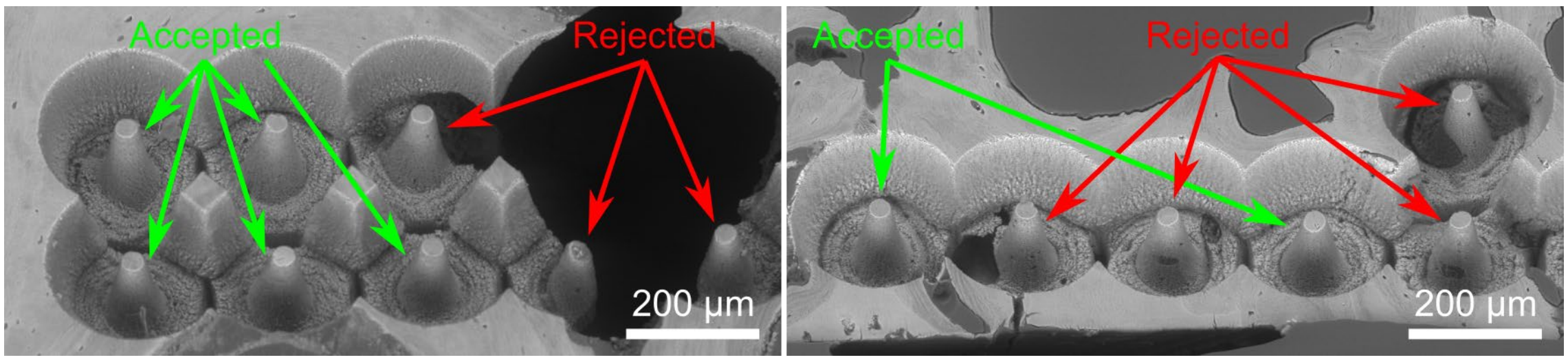

Fig. S1 – Examples of accepted and rejected micropillars. Micropillars were rejected if they had large porosity within or under the pillar, had PMMA inclusions within the pillar, or (iii) exhibited defects such as chips, cracks, or irregular geometry.

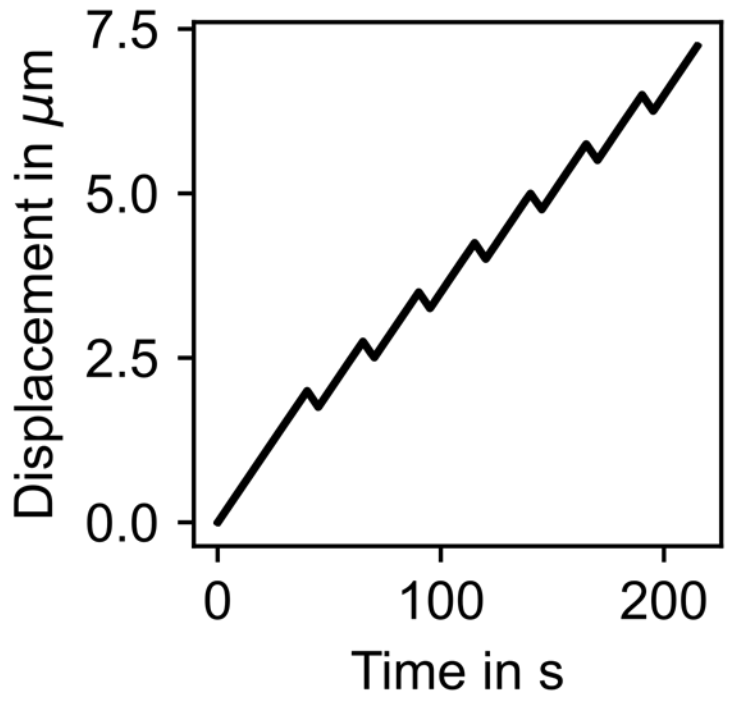

Fig. S2 - The cyclic uniaxial microindenter loading protocol.

## S1. Material model

To capture the nonlinear material behaviour of bone at the microscale, we adopt an elastoplastic material model outlined by Schwiedrzik et al. [1]. We outline the basic constitutive equations, where, scalars are denoted as standard minuscules or majuscules ($x$ or $X$), vectors are denoted as bold-face minuscule ($\boldsymbol{x}$), second order tensors denoted as bold-face font majuscules ($\boldsymbol{X}$), and forth order tensors are denoted as scripted majuscules ($\mathbb{X}$). The $\otimes$ symbol signifies the tensor product and $\boldsymbol{X} \underline{\overline{\otimes}} \boldsymbol{Y} = (\boldsymbol{X}_{ik}\boldsymbol{Y}_{jl} + \boldsymbol{X}_{il}\boldsymbol{Y}_{jk})/2$ is the symmetric product of two second order tensors. $\boldsymbol{X}_{ij} = \mathbb{Y}_{ijkl}\boldsymbol{Z}_{kl}$ is the transformation of a second-order tensor with a fourth-order tensor and ':' denotes the double contraction of two tensors.

We use an additive decomposition of the elastic and plastic strains as [2]:

$$\boldsymbol{E} = \boldsymbol{E}^e + \boldsymbol{E}^p. \qquad (S1)$$

The stress tensor, $\boldsymbol{S}$, is then given by:

$$\boldsymbol{S} = \mathbb{S}\,(\boldsymbol{E} - \boldsymbol{E}^p). \qquad (S2)$$

where, $\mathbb{S}$, is the stiffness tensor which when isotropic takes the form:

$$\mathbb{S} = \frac{E_0\nu_0}{(1+\nu_0)(1-2\nu_0)}\boldsymbol{I} \otimes \boldsymbol{I} + \frac{E_0}{1+\nu_0}\boldsymbol{I} \underline{\overline{\otimes}} \boldsymbol{I} \qquad (S3)$$

Here $\nu_0$ is the Poisson's ratio of the tissue. We use a quadric yield surface proposed by Schwiedrzik et al. [3], which is given by:

$$Y(\boldsymbol{S}, \kappa) \coloneqq \sqrt{\boldsymbol{S}: \mathbb{F}\boldsymbol{S}} + \boldsymbol{F}: \boldsymbol{S} - r(\kappa) = 0 \qquad (S4)$$

$\mathbb{F}$ and $\boldsymbol{F}$ here control the origin, shape, and orientation of the yield surface and take the form [3]:

$$\mathbb{F} = -\frac{\zeta_0}{4}\left(\frac{\sigma_0^+ + \sigma_0^-}{\sigma_0^+\sigma_0^-}\right)^2 \boldsymbol{I} \otimes \boldsymbol{I} + \frac{(1+\zeta_0)}{4}\left(\frac{\sigma_0^+ + \sigma_0^-}{\sigma_0^+\sigma_0^-}\right)^2 \boldsymbol{I} \underline{\overline{\otimes}} \boldsymbol{I} \qquad (S5)$$

and:

$$\boldsymbol{F} = \left(\frac{1}{\sigma_0^+} - \frac{1}{\sigma_0^-}\right)\boldsymbol{I}. \qquad (S6)$$

The interaction parameter $\zeta_0$, which governs the shape of the yield surface is set as 0.49 to form a rounded tipped cone that is aligned with the hydrostatic stress axis in normal stress space (Fig. S2). $\sigma_0^+$, and $\sigma_0^-$ are the uniaxial yield stress in tension and compression respectively. The yield criterion accounts for the asymmetry in tensile and compressive strength, which is evident in bone tissue at the macro- [4,5] and microscale [6] of bone. We implement a constant ratio of $\sigma_0^+ = \frac{2}{3}\sigma_0^-$ irrespective of the magnitude of $\sigma_0^-$.

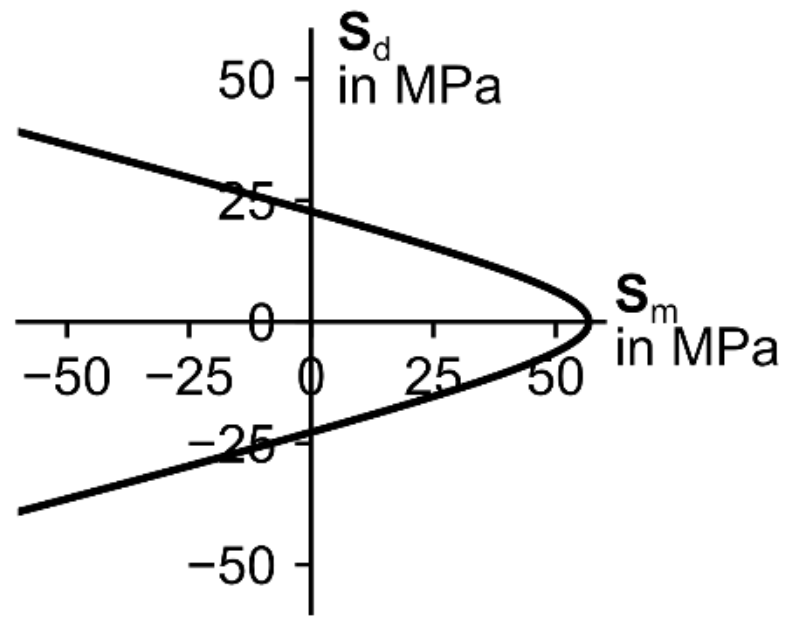

Fig. S3 – Graphical representation of the quadric yield surface. 2D illustration in the mean stress vs deviator stress plane.

$r(\kappa)$ is an isotropic hardening function which was chosen to be linear following Bayraktar et al. [7]. The hardening function, which in itself is a function of accumulated plastic strain where $\kappa = \int |\dot{\boldsymbol{E}}^p| \, \mathrm{d}t$, takes the form:

$$r(\kappa) = 1 + \frac{E_0}{\sigma_0^-} m_{py} \qquad (S7)$$

Here $m_{py}$ is coefficient which controls the gradient of the linear hardening. $E_0$, $\sigma_0^-$ and $m_{py}$ for each pillar were identified by an inverse technique by *in silico* micropillar compression. The effect of each of the parameters under uniaxial tension and compression are shown in Fig. S3.

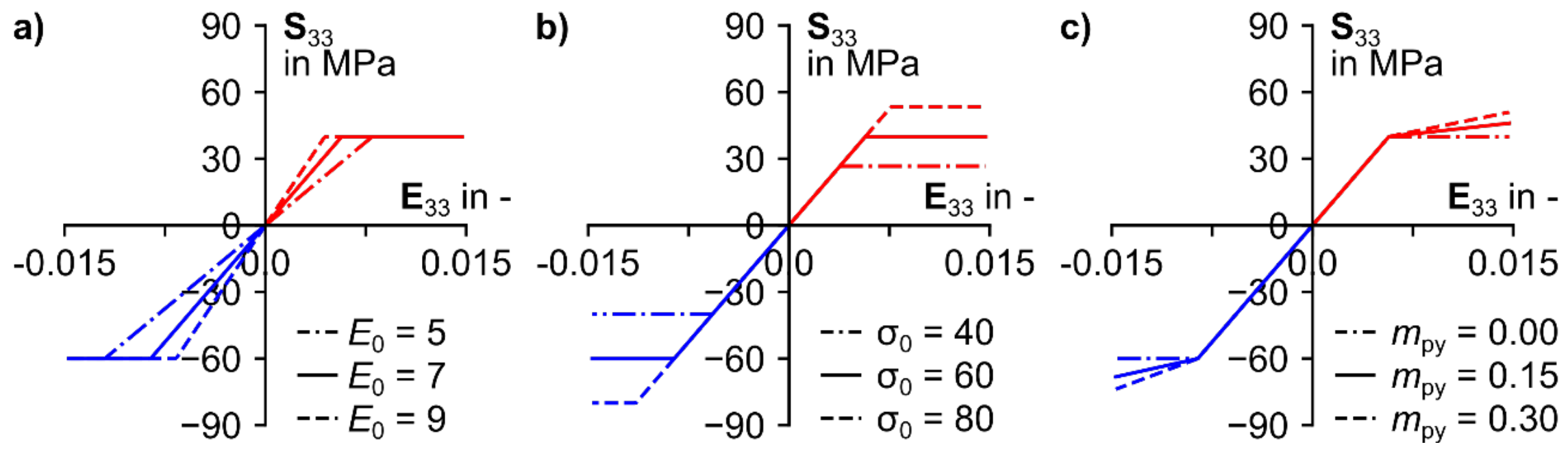

Fig. S4 – Uniaxial material behaviour. Material behaviour under uniaxial tension (red) and compression (blue). For each plot, one of the mechanical parameters is varied: a) is elastic modulus ($E_0$) in GPa, b) is compressive yield stress ($\sigma_0^-$) and c) is the post yield gradient ($m_{py}$). Note the asymmetry between tensile and compressive yield strength.

## S2. Back-calculation routine

For each pillar, $E_0$, $K_{cont}$, $\sigma_0^-$ and $m_{py}$ were derived iteratively using a secant method to minimise the relative error, $|1 - (Y_{FE}/Y_{EXP})|$, between the experimentally and numerically derived force-displacement curves. Here $Y$ is one of the target variables $E_0$, $K_{cont}$, $\sigma_0^-$ or $m_{py}$. In the following section, we outline the method for evaluating $\sigma_0^-$.

However, the iterative scheme was used for each parameter in a sequential fashion with the following order: $E_0$, $K_{cont}$, $m_{py}$, $\sigma_0^-$.

For each pillar, two initial simulations were conducted with a low $\sigma_L^-$ and high $\sigma_H^-$ compressive yield strength. The yield force for both simulations, $F_L$ and $F_H$, were computed with a 0.2% offset criterion. The iterative process then started and a trial compressive yield strength $\sigma_T^-$ was computed by:

$$\sigma_T^- = \frac{(\sigma_H^- - \sigma_L^-)(F_{EXP} - F_L)}{F_H - F_L} + \sigma_L^- \qquad (S8)$$

Here $F_{EXP}$ is the experimentally derived yield force. The model was then parameterised with $\sigma_T^-$ and resubmitted and the yield force was again evaluated now as $F_T$. The relative error $|1 - (F_T/F_{EXP})|$ was computed and if greater than a convergence threshold of $10^{-5}$, $\sigma_L^-$, $\sigma_H^-$, $F_L$ and $F_H$ were updated conditionally:

$$\sigma_L^- = \begin{cases} \sigma_L^- & \text{if} \quad F_T > F_{EXP} \\ \sigma_T^- & \text{if} \quad F_T < F_{EXP} \end{cases} \quad \text{and} \quad \sigma_H^- = \begin{cases} \sigma_T^- & \text{if} \quad F_T > F_{EXP} \\ \sigma_H^- & \text{if} \quad F_T < F_{EXP} \end{cases}$$

$$(S9)$$

$$F_L = \begin{cases} F_L & \text{if} \quad F_T > F_{EXP} \\ F_T & \text{if} \quad F_T < F_{EXP} \end{cases} \quad \text{and} \quad F_H = \begin{cases} F_T & \text{if} \quad F_T > F_{EXP} \\ F_H & \text{if} \quad F_T < F_{EXP} \end{cases}$$

This iterative process where a new trial stress is computed by Equation S8, resubmitted and the relative error evaluation repeated until the convergence threshold was satisfied. Fig. S4 shows the reduction in relative error through each iteration.

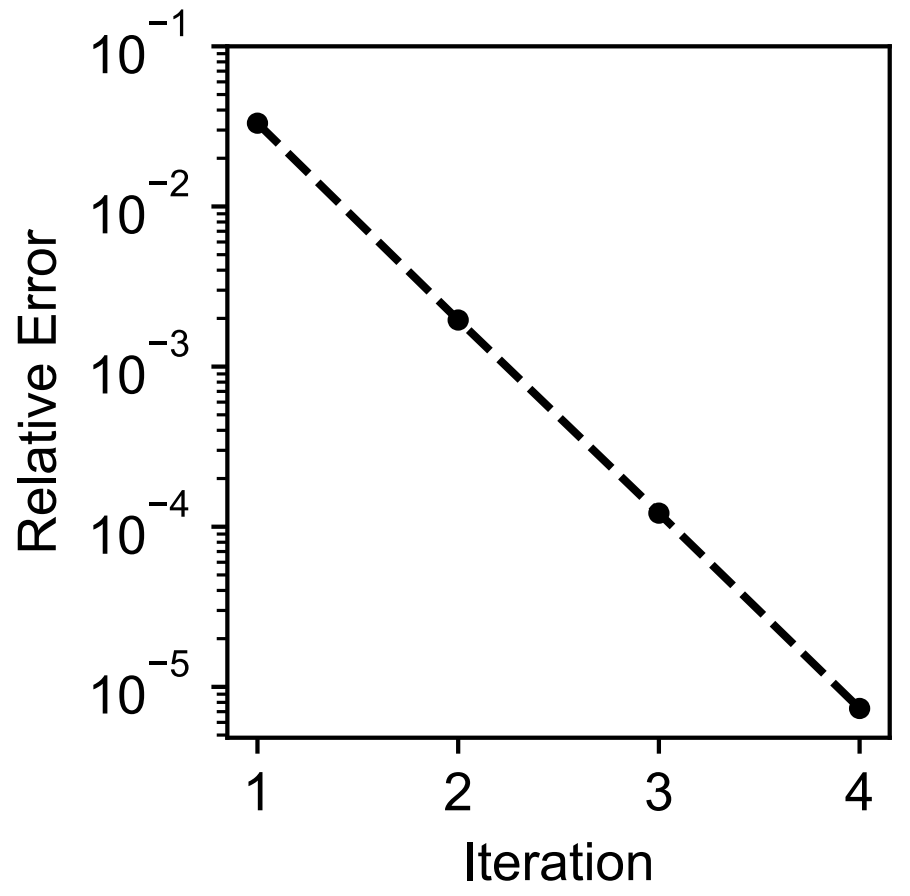

Fig. S5. – Relative error per iteration in the back-calculation scheme.

## S3. Transverse isotropy

The formulation for the transversely isotropic stiffness tensor is composed material properties and the tensor product of a unit vector, $\boldsymbol{m}_A$, in the axial direction such that $\boldsymbol{M}_A = \boldsymbol{m}_A \otimes \boldsymbol{m}_A$ and [10]:

$$
\begin{aligned}
\mathbb{S} =& \frac{E_T(E_A \nu_T + E_T \nu_A^2)}{(1+\nu_T)(E_A(1-\nu_T) - 2E_T\nu_A^2)} (\boldsymbol{I} - \boldsymbol{M}_A) \otimes (\boldsymbol{I} - \boldsymbol{M}_A) \\
&+ \frac{E_T}{1+\nu_T} (\boldsymbol{I} - \boldsymbol{M}_A) \overline{\underline{\otimes}} (\boldsymbol{I} - \boldsymbol{M}_A) \\
&+ \frac{E_A^2(1-\nu_T)}{E_A(1-\nu_T) - 2E_T\nu_A^2} \boldsymbol{M}_A \otimes \boldsymbol{M}_A \\
&+ \frac{E_T E_A \nu_A}{E_A(1-\nu_T) - 2E_T\nu_A^2} \left((\boldsymbol{I} - \boldsymbol{M}_A) \otimes \boldsymbol{M}_A + \boldsymbol{M}_A \otimes (\boldsymbol{I} - \boldsymbol{M}_A)\right) \\
&+ 2\mu_T \left((\boldsymbol{I} - \boldsymbol{M}_A) \overline{\underline{\otimes}} \boldsymbol{M}_A + \boldsymbol{M}_A \overline{\underline{\otimes}} (\boldsymbol{I} - \boldsymbol{M}_A)\right) \qquad (S10)
\end{aligned}
$$

Subscript $A$ and $T$ are axial and transverse respectively. The fourth- and second-order tensors that compose the yield criterion for a transverse isotropic case take the form [3]:

$$
\begin{aligned}
\mathbb{F} =& -\frac{\zeta_T}{4}\left(\frac{1}{\sigma_T^+} + \frac{1}{\sigma_T^-}\right)^2 (\boldsymbol{I} - \boldsymbol{M}_A) \otimes (\boldsymbol{I} - \boldsymbol{M}_A) \\
&+ \frac{\zeta_T + 1}{4}\left(\frac{1}{\sigma_T^+} + \frac{1}{\sigma_T^-}\right)^2 (\boldsymbol{I} - \boldsymbol{M}_A) \overline{\underline{\otimes}} (\boldsymbol{I} - \boldsymbol{M}_A) \\
&+ \frac{1}{4}\left(\frac{1}{\sigma_A^+} + \frac{1}{\sigma_A^-}\right)^2 \boldsymbol{M}_A \otimes \boldsymbol{M}_A \\
&- \frac{\zeta_A}{4}\left(\frac{1}{\sigma_A^+} + \frac{1}{\sigma_A^-}\right)^2 \left((\boldsymbol{I} - \boldsymbol{M}_A) \otimes \boldsymbol{M}_A + \boldsymbol{M}_A \otimes (\boldsymbol{I} - \boldsymbol{M}_A)\right) \\
&+ \frac{1}{2\tau_A^2}\left((\boldsymbol{I} - \boldsymbol{M}_A) \overline{\underline{\otimes}} \boldsymbol{M}_A + \boldsymbol{M}_A \overline{\underline{\otimes}} (\boldsymbol{I} - \boldsymbol{M}_A)\right) \qquad (S11)
\end{aligned}
$$

and:

$$
\boldsymbol{F} = \frac{1}{2}\left(\frac{1}{\sigma_T^+} - \frac{1}{\sigma_T^-}\right)(\boldsymbol{I} - \boldsymbol{M}_A) + \frac{1}{2}\left(\frac{1}{\sigma_A^+} - \frac{1}{\sigma_A^-}\right)\boldsymbol{M}_A \qquad (S12)
$$

The parameterisation of was chosen using the axial-to-transverse elasticity and strength ratios reported by Schwiedrzik et al. [11] ($E_A/E_T$ = 1.57, $\sigma_A^-/\sigma_T^-$ = 1.31).

$E_A$, $E_T$ and $\sigma_A^-$, $\sigma_T^-$ were then derived from our median (Subscript $Mdn$) micropillar $E_0$ and $\sigma_0^-$ results by:

$$E_T = \frac{2(E_{Mdn})}{2.57} \quad \text{and} \quad \sigma_T^- = \frac{2\left(\sigma_{T\ Mdn}^-\right)}{2.31} \tag{S13}$$

Table S1 lists the material properties used in the stiffness and yield criterion tensors.

Table S1 – Parameters used in the stiffness and yield criterion tensors using axial-to-transverse elasticity and strength ratios reported by Schwiedrzik et al. [11].

| **Elasticity** | | | | | **Strength** | | | | |
|---|---|---|---|---|---|---|---|---|---|
| $E_A$ | $E_T$ | $\nu_A$ | $\nu_T$ | $\mu_A$ | $\sigma_A^-$ | $\sigma_T^-$ | $\tau_A$ | $\zeta_A$ | $\zeta_T$ |
| in MPa | in MPa | in - | in - | in MPa | in MPa | in MPa | in MPa | in - | in - |
| 9579.6 | 6093.9 | 0.376 | 0.30 | 2938.6 | 58.3 | 44.6 | 23.6 | 0.641 | 0.490 |

## References

[1] J. Schwiedrzik, T. Gross, M. Bina, M. Pretterklieber, P. Zysset, D. Pahr, Experimental validation of a nonlinear $\mu$ FE model based on cohesive-frictional plasticity for trabecular bone, Numer Methods Biomed Eng 32 (2016) e02739.

[2] A.E. Green, P.M. Naghdi, A general theory of an elastic-plastic continuum, Archive for Rational Mechanics and Analysis 18 (1965) 251–281.

[3] J.J. Schwiedrzik, U. Wolfram, P.K. Zysset, A generalized anisotropic quadric yield criterion and its application to bone tissue at multiple length scales, Biomechanics and Modeling in Mechanobiology 12 (2013) 1155–1168.

[4] M.J. Mirzaali, A. Bürki, J. Schwiedrzik, P.K. Zysset, U. Wolfram, Continuum damage interactions between tension and compression in osteonal bone, Journal of the Mechanical Behavior of Biomedical Materials 49 (2015) 355–369.

[5] U. Wolfram, T. Gross, D.H. Pahr, J. Schwiedrzik, H.J. Wilke, P.K. Zysset, Fabric-based Tsai-Wu yield criteria for vertebral trabecular bone in stress and strain space, Journal of the Mechanical Behavior of Biomedical Materials 15 (2012) 218–228.

[6] D. Casari, J. Michler, P. Zysset, J. Schwiedrzik, Microtensile properties and failure mechanisms of cortical bone at the lamellar level, Acta Biomaterialia 120 (2021) 135–145.

[7] H.H. Bayraktar, E.F. Morgan, G.L. Niebur, G.E. Morris, E.K. Wong, T.M. Keaveny, Comparison of the elastic and yield properties of human femoral trabecular and cortical bone tissue, Journal of Biomechanics 37 (2004) 27–35.

[8] U. Wolfram, M. Peña Fernández, S. McPhee, E. Smith, R.J. Beck, J.D. Shephard, A. Ozel, C.S. Erskine, J. Büscher, J. Titschack, J.M. Roberts, S.J. Hennige, Multiscale mechanical consequences of ocean acidification for cold-water corals, Sci Rep 12 (2022) 8052.

[9] H. Zhang, B.E. Schuster, Q. Wei, K.T. Ramesh, The design of accurate micro-compression experiments, Scripta Materialia 54 (2006) 181–186.

[10] P.K. Zysset, A. Curnier, An alternative model for anisotropic elasticity based on fabric tensors, Mechanics of Materials 21 (1995) 243–250.

[11] J. Schwiedrzik, A. Taylor, D. Casari, U. Wolfram, P. Zysset, J. Michler, Nanoscale deformation mechanisms and yield properties of hydrated bone extracellular matrix, Acta Biomaterialia 60 (2017) 302–314.